\documentclass[11pt,a4paper]{article}
\pdfoutput=1
\usepackage{jheppub}
\usepackage{amsmath}
\usepackage{epsfig}
\usepackage{amssymb}
\usepackage{graphics}
\usepackage[active]{srcltx}
\usepackage{epstopdf}
\usepackage{shuffle}
\usepackage[utf8]{inputenc}

\usepackage[T1]{fontenc} 

\usepackage{slashed}
\usepackage{tikz}
\usepackage{adjustbox}
\usepackage{array}
\allowdisplaybreaks
\usetikzlibrary{shapes,decorations.markings}

\newcommand{\bea}{\begin{eqnarray}}
\newcommand{\eea}{\end{eqnarray}}
\newcommand{\bean}{\begin{eqnarray*}}
\newcommand{\eean}{\end{eqnarray*}}
\newcommand{\nn}{\nonumber \\}

\newcommand {\Pf}  {\text{Pf}\,}
\newcommand {\Pfp}  {\text{Pf}\,'}

\def\W #1{\widetilde{#1}}

\def\Label#1{\label{#1}%
  \smash{\hbox to0pt{\raise1ex\hbox{\tiny[#1]}\hss}}}

\def\Pf{{\mbox{Pf}}}

\def\Label{\label}
\newcommand{\Sl}{\sum\limits}

\title{Gauge invariance induced relations and the equivalence between distinct approaches to  NLSM amplitudes}

\author[a,b]{Yi-Jian Du}
\author[c,d]{,\,Yong Zhang}

\affiliation[a]{Center for Theoretical Physics, School of Physics and Technology,
Wuhan University, \\
No.299 Bayi Road, Wuhan 430072, China}
\affiliation[b]{Suzhou Institute of Wuhan University,\\
No.377 Linquan Street, Suzhou, 215123, China}

\affiliation[c]{CAS Key Laboratory of Theoretical Physics, Institute of Theoretical Physics, Chinese Academy
of Sciences,\\
 Beijing, 100190, China}
\affiliation[d]{
University of Chinese Academy of Sciences, No.19A Yuquan Road, Beijing 100049, China}

\emailAdd{yijian.du@whu.edu.cn, yongzhang@itp.ac.cn}

\date{\today}
\abstract{In this paper, we derive generalized  Bern-Carrasco-Johansson (BCJ) relations for color-ordered Yang-Mills amplitudes by imposing gauge invariance conditions and dimensional reduction appropriately on the new discovered graphic expansion of Einstein-Yang-Mills amplitudes.
These relations are also satisfied by color-ordered amplitudes in other theories such as bi-scalar theory and nonlinear sigma model (NLSM). As an application of the gauge invariance induced relations, we further prove that the three types of BCJ numerators in {NLSM}, which are derived from Feynman rules, Abelian Z-theory and Cachazo-He-Yuan (CHY) formula respectively, produce the same total amplitudes. In other words, the three distinct approaches to NLSM amplitudes are equivalent to each other.
}
\keywords{Amplitude Relation, BCJ Numerator, Gauge Invariance}

\begin{document}
\maketitle \flushbottom

\section{Introduction}\label{SecIntroduction}
Color-kinematic duality (BCJ duality), which was suggested by Bern Carrasco and Johansson \cite{Bern:2008qj, Bern:2010ue}, provides a deep insight into the study of scattering amplitudes. According to BCJ duality, full color-dressed Yang-Mills amplitudes are expressed by summing over trivalent (Feynman-like) diagrams, each of which is associated with a color factor and a kinematic factor (BCJ numerator) sharing the same algebraic properties ({\it i.e.}, antisymmetry and Jacobi identity). Once the color factors are replaced by BCJ numerators of another copy Yang-Mills amplitude, we obtain a gravity amplitude.

A significant consequence of BCJ duality is that tree-level color-ordered Yang-Mills amplitudes satisfy BCJ relations where the coefficients for amplitudes are functions of Mandelstam variables. Together with the earlier proposed Kleiss-Kuijf \cite{Kleiss:1988ne} (KK) relations, BCJ relations reduce the number of independent color-ordered Yang-Mills amplitudes to $(n-3)!$ (see the field theory proofs \cite{Feng:2010my,Chen:2011jxa} and string theory approaches \cite{BjerrumBohr:2009rd,Stieberger:2009hq}). Though BCJ relations are first discovered in Yang-Mills theory, they actually hold for amplitudes in many other theories including:  bi-scalar theory, NLSM \cite{Chen:2013fya}, which can be uniformly described in the framework of CHY formulation \cite{Cachazo:2013gna,Cachazo:2013hca,Cachazo:2013iea,Cachazo:2014xea}. It was pointed out that fundamental BCJ relation can be regarded as the most elementary one since the minimal basis \cite{Feng:2010my} and a set of more general BCJ relations \cite{BjerrumBohr:2009rd,Chen:2011jxa} are generated by them \cite{Ma:2011um}. Nevertheless, in some situations, one may encounter BCJ relations which have much more complicated forms than knowns ones. Such relations can be neither directly understood as a result of fundamental relations nor straightforwardly proved by Britto-Cachazo-Feng-Witten \cite{Britto:2004ap,Britto:2005fq} recursion or CHY formula. Therefore, a new approach to nontrivial BCJ relations is required.

Apart from the BCJ relations for amplitudes, the construction of BCJ numerators in various theories is also an important direction. In NLSM, there are three distinct constructions of BCJ numerators, all of which are polynomial functions of Mandelstam variables. (i) A construction based on off-shell extended BCJ relation (see \cite{Chen:2013fya}) was suggested by Fu and one of the current authors \cite{Du:2016tbc} (DF). In DF approach, {the set of half-ladder numerators with the first and the last lines fixed (which serves as a basis of BCJ numerators)} are expressed by proper combinations of momentum kernels \cite{Kawai:1985xq,Bern:1998ug,BjerrumBohr:2010ta,BjerrumBohr:2010zb,BjerrumBohr:2010yc,BjerrumBohr:2010hn}. Since the off-shell extended BCJ relation \cite{Chen:2013fya} was proved by the use of Berends-Giele recursion (Feyman rules), the DF type BCJ numerators can be essentially regarded as a result of Feyman rules. (ii) {A much more compact} construction of BCJ numerators in NLSM, {which was based on Abelian Z theory}, was provided by Carrasco, Mafra and  Schlotterer (CMS) \cite{Carrasco:2016ldy}. A half ladder numerator of CMS type is {elegantly expressed} by only one momentum kernel. (iii) In a  more recent work \cite{Du:2017kpo}, a graphic approach to polynomial BCJ numerators (DT type numerator) in NLSM, which was based on CHY formula was proposed by Teng and one current author. All the three distinct constructions given above must produce the same scattering amplitudes in NLSM, but this equivalence is still not proven {explicitly}.

In this paper, we derive highly nontrivial generalized BCJ relations (gauge invariance induced relations) by imposing gauge invariance conditions and CHY-inspired dimensional reduction on the recent discovered graphic expansion of color-ordered Einstein-Yang-Mills (EYM) amplitudes \cite{Du:2017kpo}. Expansion of EYM amplitudes was first proposed in \cite{Stieberger:2016lng} and further studied in \cite{Nandan:2016pya,delaCruz:2016gnm,Schlotterer:2016cxa,Fu:2017uzt,Chiodaroli:2017ngp,Teng:2017tbo,Du:2017kpo,Du:2017gnh}. In the series work \cite{Fu:2017uzt,Teng:2017tbo,Du:2017kpo,Du:2017gnh}, general recursive expansion for all tree-level EYM amplitudes and the graphic expansion of EYM amplitudes in terms of pure Yang-Mills ones were established. When gauge invariance condition for the so-called fiducial graviton is imposed, the recursive expansion of EYM amplitudes induces relations between those amplitudes with fewer gravitons. Equivalently, when the graphic expansion \cite{Du:2017kpo} is considered, such gauge invariance induced relation implies a relation between color ordered Yang-Mills amplitudes whose coefficients are functions of both momenta and polarizations. To induce amplitude relations where all coefficients are functions of Mandelstam variables, one should convert all polarizations in the coefficients into momenta.
In the current paper, we propose gauge invariance induced relations based on the following two crucial observations: (i) One can impose the gauge invariance conditions for several gravitons simultaneously. (ii) The gauge invariance conditions are independent of dimensions. With these two {critical observations} in hand and inspired by the dimensional reduction in CHY formula \cite{Cachazo:2014xea}, we define $(d+d)$-dimensional polarizations and momenta whose { nonzero} components are expressed by only $d$-dimensional momenta. Imposing the gauge invariance in $(d+d)$ dimensions on the graphic expansion \cite{Du:2017kpo} of {single-trace} EYM amplitudes, we naturally induce nontrivial amplitude relations where all coefficients are polynomials of Mandelstam variables (in $d$ dimensions). In the framework of CHY formula, such relations become nontrivial relations between Parke-Taylor factors.
As a consequence, the gauge invariance induced relations hold for not only color-ordered Yang-Mills amplitudes but also {color-ordered amplitudes in other theories such as  bi-scalar theory and NLSM.}

An interesting application of our gauge invariance induced relation is the proof of equivalence between different approaches to NLSM amplitudes.
Full color-dressed NLSM amplitudes can be spanned in terms of bi-scalar amplitudes via dual Del Duca-Dixon-Maltoni (DDM) \cite{DelDuca:1999rs} decomposition (The dual DDM decomposition for Yang-Mills amplitudes are given in \cite{Kiermaier,BjerrumBohr:2010hn,Bern:2010yg,Mafra:2011kj,Du:2011js,Fu:2012uy,Cachazo:2013iea,Fu:2013qna, Du:2013sha, Fu:2014pya}, for NLSM amplitudes are provided in \cite{Chen:2013fya,Du:2016tbc,Carrasco:2016ldy,Du:2017kpo}), in which the coefficients are half-ladder BCJ numerators {with fixing the first and the last lines}. Although the three distinct approaches: Feyman rules, Abelian Z theory and CHY formula provide different types of half-ladder BCJ numerators, they must produce the same NLSM amplitudes through the dual DDM decomposition. This equivalence condition then requires nontrivial relations between 
color-ordered bi-scalar amplitudes.
By using the gauge invariance induced relations and defining partial momentum kernel, we prove that the three distinct constructions of BCJ numerators produce the same NLSM amplitudes precisely. In other words, the equivalence between the three different approaches to NLSM amplitudes is {explicitly} proven.
The relation between main results of this paper is provided as
\bea
  \begin{array}{c}
    \text{gauge invariance}\\
    +\\
    \text{dimensional reduction}\\
  \end{array}
\Rightarrow\text{generalized BCJ \eqref{Eq:NewGaugeIDAmp1}}\Rightarrow\begin{array}{ccc}
&\text{relation \eqref{Eq:GenEquiv1}}&\Rightarrow\text{equivalence between CMS $\&$ DT}\\
  \nearrow & &  \\
   \searrow & & \\
   &\text{relation \eqref{Eq:GenEquiv2}} &\Rightarrow\text{equivalence between DF $\&$ CMS}\\
  \end{array}.\nonumber
\eea

The structure of this paper is given as follows. In section \ref{Sec:CHY}, we provide a review of the background knowledge including CHY formula, the recursive expansion and the graphic expansion of EYM amplitudes. In section \ref{BCJ-Gauge}, we induce generalized BCJ relations by combining gauge invariance conditions and dimensional reduction. Partial momentum kernel, which is important for the discussions in this paper, is introduced in section \ref{BCJ-Gauge}. A review of the three distinct constructions of BCJ numerators in NLSM is provided in section \ref{Sec:NumeratorsNLSM1}. In section \ref{Sec:DTCMS}, we prove the equivalence between CMS type and DT type numerators by inducing identities expressed by partial momentum kernel. The proof of equivalence between DF type and CMS type numerators is given in section \ref{Sec:DFCMS}. We summarized this paper in section \ref{Sec:Conclusions}. Complicated graphs and proofs are included by appendices.

\section{A review of CHY formula and the expansion of EYM amplitudes}\label{Sec:CHY}
In this section, we review the CHY formula \cite{Cachazo:2013gna,Cachazo:2013hca,Cachazo:2013iea,Cachazo:2014nsa,Cachazo:2014xea} for various theories and the recursive/graphic expansion of EYM amplitudes which will be used in the coming sections.
\subsection{CHY formula}
CHY formula expresses a tree level on-shell amplitude with $n$ massless particles by integration over $n$ scattering variables $z_i$
\bea
A=\int d\Omega_{\text{CHY}} \mathcal{I}_L \mathcal{I}_R,~~\Label{Eq:CHY}
\eea
where  $d\Omega_{\text{CHY}}$ is M\"{o}bius invariant measure which contains the condition that scattering variables satisfy the following scattering equations
\bea
\Sl_{j\neq i}{k_i\cdot k_j\over z_i-z_j}=0,~~~\text(i=1,\dots,n).\Label{Eq:SE}
\eea
Here $k_i$ denotes the momenta of the particle $i$. The integrand $\mathcal{I}_L \mathcal{I}_R$ in \eqref{Eq:CHY} relies on theories. An important feature is that the CHY formula is independent of dimensions.
\subsubsection*{The CHY integrand for BS, CS, YM, EYM and GR amplitudes}
The CHY integrands for color-ordered bi-scalar {(BS)}, Yang-Mills {(YM)}, single-trace EYM amplitudes {(EYM)} as well as gravity {(GR)} amplitudes are given by\footnote{ The total signs follows from the paper \cite{Teng:2017tbo}.}
\bea
\mathcal{I}^{\text{BS}}_L(\pmb{\sigma}_{1,n})&=&(-1)^{{(n+1)(n+2)\over 2}}\text{PT}(\pmb{\sigma}_{1,n}),~~~~~~~~\mathcal{I}^{\text{BS}}_R(\pmb{\rho}_{1,n})=(-1)^{{(n+1)(n+2)\over 2}}\text{PT}(\pmb{\rho}_{1,n})~~\Label{Eq:CHYBSintegrand}
\\
\mathcal{I}^{\text{YM}}_L(\pmb{\sigma}_{1,n})&=&(-1)^{{(n+1)(n+2)\over 2}}\text{PT}(\pmb{\sigma}_{1,n}),~~~~~~~~~~~~~~~~~~~~~\mathcal{I}^{\text{YM}}_R\,=\Pf\,'[\Psi]~~\Label{Eq:CHYYMintegrand}\\
\mathcal{I}^{\text{EYM}}_L(\pmb{\sigma}_{1,r})&=&(-1)^{{(n+1)(n+2)+s(s+1)\over 2}}\text{PT}(\pmb{\sigma}_{1,r})\Pf[\Psi_{\mathsf{H}}],~~~~\mathcal{I}^{\text{EYM}}_R=\Pf\,'[\Psi]~~\Label{Eq:CHYEYMintegrand}\\
\mathcal{I}^{\text{GR}}_R&=&\Pf\,'[\Psi],~~~~~~~~~~~~~~~~~~~~~~~~~~~~~~~~~~~~~~~~~~~\mathcal{I}^{\text{GR}}_R\,=\Pf\,'[\Psi].~~\Label{Eq:CHYGRintegrand}
\eea
In \eqref{Eq:CHYBSintegrand} and \eqref{Eq:CHYYMintegrand}, the boldface Greek letters $\pmb{\sigma}_{1,n}$ and $\pmb{\rho}_{1,n}$ denote permutations  of all $n$ external particles $1, 2,\dots, n$. The Parke-Taylor factor $\text{PT}(\pmb{\sigma}_{1,n})$ is defined by
\bea
\text{PT}(\pmb{\sigma}_{1,n})={1\over z_{\sigma(1)\sigma(2)}z_{\sigma(2)\sigma(3)}\dots z_{\sigma(n)\sigma(1)}},~~~~z_{ij}\equiv z_i-z_j.
\eea
The reduced Pfaffian $\Pf\,'[\Psi]$ in \eqref{Eq:CHYYMintegrand}, \eqref{Eq:CHYEYMintegrand} and  \eqref{Eq:CHYGRintegrand} is given by
\bea
\Pf\,'\left[\Psi\right]\equiv{(-1)^{i+j}\over z_{ij}}\Pf\left[\Psi^{i,j}_{i,j}\right],~~~~~~~~~~~~~\Psi=\left(\begin{array}{cc}
    {A} & -{C}^{T} \\
   {C} &{B}
    \end{array}\right)\,,
\eea
where $\Psi^{i,j}_{i,j}$ means that  the $i$, $j$-th ($1\leq i,j\leq n$) rows and columns are removed.
Building blocks of the $2n\times 2n$-skew matrix $\Psi$ are
\begin{equation}
   {A}_{ab}=\Biggl\{\begin{array}{cc}
                                                {k_a\cdot k_b\over z_{ab}} &~~a\neq b \\
                                                0 &~~a=b
                                            \end{array}~~~~{B}_{ab}=\Biggl\{\begin{array}{cc}
                                                {\epsilon_a\cdot \epsilon_b\over z_{ab}} &~~a\neq b \\
                                                0 &~~a=b
                                              \end{array}
                                              ~~~~{C}_{ab}=\Biggl\{\begin{array}{cc}
                                                {\epsilon_a\cdot k_b\over z_{ab}} &~~a\neq b \\
                                                -\Sl_{c\neq a}{\epsilon_a\cdot k_c\over z_{ac}} &~~a=b
                                              \end{array},\Label{Eq:CHYBlocks}
\end{equation}
in which $k_a$ and $\epsilon_a$ are momentum and polarization of the particle $a$. In the CHY expression of single-trace EYM amplitude \eqref{Eq:CHYEYMintegrand}, $\text{PT}(\pmb{\sigma}_{1,r})$ denotes the Parke-Taylor factor for $r$ gluons with the order $\sigma(1),\sigma(2),\dots,\sigma(r)$. The matrix $\Psi_{\mathsf{H}}$  is the one obtained by removing those rows and columns with respect to gluons in $\Psi$.

\subsubsection*{The CHY integrand for NLSM amplitudes}

The CHY integrands for color-ordered NLSM amplitudes are obtained by dimensional reduction strategy \cite{Cachazo:2014xea}.
In particular,  $\mathcal{I}_L^{\text{NLSM}}$ has the same expression with $\mathcal{I}_L^{\text{YM}}$, while $\mathcal{I}_R^{\text{NLSM}}$ is obtained by extending  $\mathcal{I}_R^{\text{YM}}$ to $(d+d+d)$-dimensions and defining  momenta and polarizations as follows:
\begin{align}
    \mathcal{K}_{a}=(k_a;0;0)& &\mathcal{E}_{a}=\left\{\begin{array}{>{\displaystyle}l @{\hspace{1.5em}} >{\displaystyle}l}
    (0;0;\frac{\epsilon_a}{\sqrt{k_1\cdot k_n}}) & a=1\text{ and }n \\
    (0;\epsilon_a;0) & a=2\ldots n-1
    \end{array}\right.\,.\Label{Eq:CHYreduction}
\end{align}
The matrix $\Psi^{(d+d+d)}$ is thus written as
\begin{equation}
    \Psi^{(d+d+d)}=\left(\begin{array}{cc}
    \mathbb{A} & -\mathbb{C}^{T} \\
   \mathbb{C}& \mathbb{B}
    \end{array}\right)\,,
\end{equation}
where the $\mathbb{A}$, $\mathbb{B}$, $\mathbb{C}$ are defined via replacing the polarizations and momenta in  \eqref{Eq:CHYBlocks} by the $(d+d+d)$-dimensional ones $\mathcal{E}$ and $\mathcal{K}$ correspondingly.
With the explicit components given in \eqref{Eq:CHYreduction}, we immediately arrive $\mathbb{C}=0$, $\mathbb{A}=A$ and $\mathbb{B}=B$. As a consequence, the reduced Pfaffian $\Pf\,'\left[\Psi^{(d+d+d)}\right]$ is factorized into:
\begin{equation}
    \Pfp\big[\Psi^{(d+d+d)}\big]=\Pfp(A)\Pf({B})=\frac{(-1)^{n+1}}{\sigma_{1n}}\frac{\epsilon_1\cdot \epsilon_n }{{k_1\cdot k_n}}\,\Pfp(A)\,\Pf(B_{1,n}^{1,n})\,.
\end{equation}
By a further replacement $\epsilon_a\rightarrow k_a$, we reduce $\Pfp\big[\Psi^{(d+d+d)}\big]$ to the final expression of the NLSM integrand $\mathcal{I}_{R}^{\text{NLSM}}$
\begin{equation}
\label{eq:DR}
    \left.\Pfp\big[\Psi^{(d+d+d)}\big]\right|_{\epsilon_a\rightarrow k_a}=\left[\Pfp(A)\right]^{2}=\mathcal{I}_R^{\text{NLSM}}\,.
\end{equation}

 \textit{{To sum up, NLSM amplitudes are obtained by performing the following replacements on  Yang-Mills amplitudes}}
\begin{align}
\label{eq:NLSMreplacement}
    &\epsilon_a\cdot k_b\;\rightarrow\;0 \nonumber\\ &\epsilon_a\cdot\epsilon_b\;\rightarrow\;\left\{\begin{array}{>{\displaystyle}l @{\hspace{1.5em}} >{\displaystyle}l}
    k_a\cdot k_b & \{a,b\}\subset\{2\ldots n-1\}\\
    1 & \{a,b\}=\{1,n\}\\
    0 & a\in\{1,n\}\text{ and }b\in\{2\ldots n-1\}\,\text{, or vice versa}
    \end{array}\right .
\end{align}

\subsection{Expansions of EYM amplitudes}\label{Sec:Expansion}
Tree level color-ordered EYM amplitude can be expressed recursively by ones with fewer gravitons and/or fewer traces. One can repeat this expansion until all amplitudes become pure Yang-Mills ones, then the expansion coefficients are constructed by graphic rules. Now we review the expansions of single-trace EYM amplitudes. The expansions of multi-trace amplitudes can be found in \cite{Du:2017gnh}.

\subsubsection*{The recursive expansion of single-trace EYM amplitudes}
Single-trace EYM amplitude $A(1,2,\dots,r\Vert\,\mathsf{H})$ with $r$ gluons and $s$ gravitons was shown to satisfy the following recursive expansion \cite{Fu:2017uzt}
\bea
A(1,2,\dots,r\Vert\,\mathsf{H})&=&\Sl_{\pmb{h}\vert\,\W{\mathsf{h}}}C_{h_i}(\pmb{h}) A(1,\{2,\dots,r-1\}\shuffle\{\pmb{h},h_i\},r\Vert\,\W{\mathsf{h}}).\Label{Eq:RecursiveExpansion}
\eea
In the above equation, we choose a fiducial graviton $h_i\in\mathsf{H}$. The summation notation stands for the sum over all possible splittings of the graviton set $\mathsf{H}\setminus {h_i}\to \pmb{h}\vert\,\W{\mathsf{h}}$ and sum over all permutations of elements in $\pmb{h}$ for a given splitting. For example, if we have three gravitons $\mathsf{H}=\{h_1,h_2,h_3\}$ and choose $h_3$ as the fiducial graviton, then $\pmb{h}\vert\,\W{\mathsf{h}}$ implies the following five terms
\bea
~~\mathsf{H}\setminus \{h_3\}&\to& \emptyset\,|\,\{h_1,h_2\};\nn
~~\mathsf{H}\setminus \{h_3\}&\to& \{h_1\}\,|\,\{h_2\};~~\mathsf{H}\setminus \{h_3\}\to \{h_2\}\,|\,\{h_1\};\nn
~~\mathsf{H}\setminus \{h_3\}&\to& \{h_1,h_2\}\,|\,\emptyset;~~\mathsf{H}\setminus \{h_3\}\to \{h_2,h_1\}\,|\,\emptyset. \Label{Eq:Terms3GRExample}
\eea
 Assuming the permutation of elements of given $\pmb{h}$ is $\{i_1,i_2,\dots,i_j\}$, the coefficient $C_{h_i}(\pmb{h})$ is defined by
\bea
C_{h_i}(\pmb{h}_1)\equiv \epsilon_{h_i}\cdot F_{i_j}\cdot F_{i_{j-1}}\cdot\dots \cdot F_{i_1}\cdot Y_{i_1},\Label{Eq:RecExpCoefficient}
\eea
where $F_a^{\mu\nu}$ is the linearized field strength of particle $a$
\bea
F_a^{\mu\nu}\equiv k_a^{\mu}\epsilon_a^{\nu}-\epsilon_a^{\mu} k_a^{\nu}
\eea
and $Y_{i_1}$ denotes the sum of all momenta of gluons in the original gluon set which appear on the left hand side of $i_1$.
An explicit example is given by the expansion of the single-trace EYM amplitude $A(1,2,\dots,r\Vert\,h_1,h_2,h_3)$ with $r$ gluons and three gravitons.
By choosing $h_3$ as the fiducial graviton and summing over the five terms in \eqref{Eq:Terms3GRExample}, we finally express the single-trace EYM amplitude with three gravitons by those amplitudes  with two, one and no graviton:
\bea
A(1,2,\dots,r\Vert\,h_1,h_2,h_3)&=&(\epsilon_{h_3}\cdot Y_{h_3})A(1,\{2,\dots,r-1\}\shuffle\{h_3\},r\Vert\,h_1,h_2)\nn
&+&(\epsilon_{h_3}\cdot F_{h_1}\cdot Y_{h_1})A(1,\{2,\dots,r-1\}\shuffle\{h_1,h_3\},r\Vert\,h_2)\nn
&+&(\epsilon_{h_3}\cdot F_{h_2}\cdot Y_{h_2})A(1,\{2,\dots,r-1\}\shuffle\{h_2,h_3\},r\Vert\,h_1)\nn
&+&(\epsilon_{h_3}\cdot F_{h_1}\cdot F_{h_2}\cdot Y_{h_2})A(1,\{2,\dots,r-1\}\shuffle\{h_2,h_1,h_3\},r)\nn
&+&(\epsilon_{h_3}\cdot F_{h_2}\cdot F_{h_1}\cdot Y_{h_1})A(1,\{2,\dots,r-1\}\shuffle\{h_1,h_2,h_3\},r).
\eea
%
\subsubsection*{Graphic rule for the pure Yang-Mills expansion of single-trace EYM amplitudes}
Applying the recursive expansion \eqref{Eq:RecursiveExpansion} repeatedly until there is no graviton remaining in the graviton set, we finally expand the single-trace EYM amplitude in terms of color-ordered Yang-Mills amplitudes
\bea
A(1,2,\dots,r\Vert{\mathsf{H}})&=&\Sl_{\pmb{\sigma}\in\{2,\dots,r-1\}\shuffle\,\text{perms}\,{\mathsf{H}}} \mathcal{C}(1,\pmb{\sigma},r)A(1,\pmb{\sigma},r).\Label{Eq:PureYMExpansion}
\eea
Here, we summed over all possible permutations obtained by merging together the original gluon set $\{2,\dots,r-1\}$ and the set of gluons (`half gravitons') which come from the graviton set $\mathsf{H}$. The relative order of gluons should be preserved, while the `perms' under the summation notation means that all possible relative orders of elements in $\mathsf{H}$ should be considered. Given order $\pmb{\sigma}$, the full coefficient $\mathcal{C}(1,\pmb{\sigma},r)$ can be determined by the following graphic rule\footnote{The interpretation of this rule is  different from that given in \cite{Du:2017kpo}, for the convenience of discussions in the coming sections.}.
\\

~~~~~~~~~~~~~~~~~~~~~~\textbf{\emph{Graphic rule for the expansion of EYM amplitudes:}}
\begin{itemize}
\item [(1)] Define a reference order $\pmb{\rho}$ of gravitons, then all gravitons are arranged into an ordered set
\bea
\mathsf{R}=\{h_{\rho(1)},h_{\rho(2)},\dots,h_{\rho(s)}\}.~~\Label{Eq:ReferenceOrder}
\eea
\item [(2)] Pick the last graviton $h_{\rho(s)}$ in the ordered set $\mathsf{R}$, an arbitrary gluon $l\in \{1,2,\dots,r-1\}$ (noting that the gluon $r$ is not considered here) as well as gravitons  $h_{i_1},h_{i_2}, \dots,h_{i_j}\in \mathsf{H}$ s.t. the relative order of them in $\pmb{\sigma}$ satisfies\footnote{In this paper, element in the $i$-th position of permutation $\pmb{\sigma}$ is denoted by $\sigma(i)$. If $\sigma(i)=a$, the position of $a$ in this permutation is denoted by $i=\sigma^{-1}(a)$. } $\sigma^{-1}(l)< \sigma^{-1}(h_{i_1})< \sigma^{-1}(h_{i_2})<\dots \sigma^{-1}(h_{i_j})< \sigma^{-1}(h_{\rho(s)})$. Now consider each particle in the set $\{l,h_{i_1},h_{i_2}, \dots,h_{i_j},h_{\rho(s)}\}$ as a node, we define a \emph{chain}  starting from the node $h_{\rho(s)}$ and ending at the node $l$. The graviton $h_{\rho(s)}$ here is mentioned as a \emph{the starting point of this chain}, while the gluon $l$ is mentioned as a \emph{root}. All other gravitons on this chain are mentioned as \emph{internal nodes of this chain}.   The factor associated to this chain is
\bea
\epsilon_{h_{\rho(s)}}\cdot F_{h_{i_j}}\cdot F_{h_{i_{j-1}}}\cdot \dots \cdot F_{h_{i_1}}\cdot k_l.
\eea
Remove $h_{i_1}$, $h_{i_2}$, ..., $h_{i_j}$, $h_{\rho(s)}$ from the ordered set $\mathsf{R}$ and redefine $\mathsf{R}$
\bea
\mathsf{R}\to\mathsf{R}\,'=\mathsf{R}\setminus \{h_{i_1},h_{i_2}, ...,h_{i_j},h_{\rho(s)}\}.
\eea
\item [(3)] Picking $l'\in \{1,2,\dots,r-1\}\cup\{h_{i_1},h_{i_2}, ...,h_{i_j},h_{\rho(s)}\}$, the last element $h_{\rho'(s')}$ in $\mathsf{R}\,'$ as well as gravitons $h_{i'_1}$, $h_{i'_2}$, ..., $h_{i'_{j'}}$ {in $\mathsf{R}\,'$} s.t.,  $\sigma^{-1}(l')<\sigma^{-1}(h_{i_1'})<\sigma^{-1}(h_{i_2'})<\dots<\sigma^{-1}(h_{i_{j'}'})<\sigma^{-1}(h_{\rho'(s')})$, we define a chain $\{l',h_{i_1'},h_{i_2'},\dots,h_{i_{j'}'},h_{\rho(s')}\}$ starting from $h_{\rho(s')}$ and ending at $l'$. This chain is associated with a factor
    \bea
    \epsilon_{h_{\rho'(s')}}\cdot F_{h_{i'_{j'}}}\cdot F_{h_{i'_{{j'-1}}}}\cdot \dots \cdot F_{h_{i_{1}'}}\cdot k_{l'}.
    \eea
    Remove $h_{i_1'}$,$h_{i_2'}$, ..., $h_{i_{j'}'}$, $h_{\rho'(s')}$ from $\mathsf{R}\,'$ and redefine $\mathsf{R}\to\mathsf{R}\,''=\mathsf{R}'\setminus\{h_{i_1'},h_{i_2'},\dots, h_{i_{j'}'},h_{\rho'(s')}\}$.
\item [(4)] Repeating the above steps until the ordered set $\mathsf{R}$ becomes empty, we get a graph (`forest') with gluons as roots of trees \footnote{Note that a starting point of a chain is not necessary a  leaf of a tree.}. For a given graph $\mathcal{F}$, the product of  the factors accompanied to all chains produces a term $\mathcal{C}^{[\mathcal{F}]}(\pmb{\sigma})$ in the coefficient $\mathcal{C}(1,\pmb{\sigma},r)$ in \eqref{Eq:PureYMExpansion}. Thus the final expression of $\mathcal{C}(1,\pmb{\sigma},r)$ is given by summing over all possible graphs defined above
    \bea
    \mathcal{C}(1,\pmb{\sigma},r)=\Sl_{\mathcal{F}\in\{\text{Graphs}\}}\mathcal{C}^{[\mathcal{F}]}(1,\pmb{\sigma},r).\Label{Eq:Coefficients}
    \eea
\end{itemize}

\subsubsection*{The expansions of Pfaffians in the CHY formula of single-trace EYM amplitudes}
It is worth closing this section by translating the expansions \eqref{Eq:RecursiveExpansion}, \eqref{Eq:PureYMExpansion} of EYM amplitudes into the language of CHY formulation (see \cite{Teng:2017tbo}). In CHY formulation, the recursive expansion \eqref{Eq:RecursiveExpansion} reflects
\bea(-1)^{s(s+1)\over 2}\text{PT}(1,2,\dots,r)\Pf\left[\Psi_{\mathsf{H}}\right]
&=&\Sl_{\pmb{h}\vert\,\W{\mathsf{h}}}(-1)^{|\W{\mathsf{h}}|(|\W{\mathsf{h}}|+1)\over 2}C_{h_1}(\pmb{h})\text{PT}(1,\{2,\dots,r-1\}\shuffle\{\pmb{h},h_1\},r)\Pf\left[\Psi_{\W{\mathsf{h}}}\right] ,\Label{Eq:RecPfaffian}\nn
\eea
where $r$ and $s$ are the numbers of gluons and gravitons respectively, $|\W{\mathsf{h}}|$ denotes the number of elements in the set $\W{\mathsf{h}}$.
The pure Yang-Mills expansion \eqref{Eq:PureYMExpansion} implies
\bea
(-1)^{s(s+1)\over 2}\text{PT}(1,2,\dots,r)\Pf\left[\Psi_{\mathsf{H}}\right]=\Sl_{\pmb{\sigma}\in\{2,\dots,r-1\}\shuffle\,\text{perms}\,{\mathsf{H}}} \mathcal{C}(1,\pmb{\sigma},r) \text{PT}(1,\pmb{\sigma},r).\Label{Eq:GraphicPfaffian}
\eea
The expansion coefficients $C_{h_1}(\pmb{h})$ and $\mathcal{C}(1,\pmb{\sigma},r)$ in  \eqref{Eq:RecPfaffian} and \eqref{Eq:GraphicPfaffian} are given by \eqref{Eq:RecExpCoefficient} and \eqref{Eq:Coefficients} respectively. We emphasize that the relations \eqref{Eq:RecPfaffian} and \eqref{Eq:GraphicPfaffian} hold for arbitrary dimensions.
\section{Gauge invariance induced relations}
\label{BCJ-Gauge}

In this section, we induce nontrivial generalized BCJ relations for color-ordered Yang-Mills amplitudes (also bi-scalar amplitudes and color-ordered NLSM amplitudes) by combining gauge invariance conditions with CHY inspired dimensional reductions. The coefficients of amplitudes in the gauge invariance induced relations  are polynomials of Mandelstam variables.

\subsection{Inducing generalized BCJ relations by gauge invariance and dimensional reduction}
In the pure Yang-Mills expansion \eqref{Eq:PureYMExpansion} of EYM amplitude $A(1,2,\dots,r\Vert\,\mathsf{H})$, each term $\mathcal{C}^{[\mathcal{F}]}(1,\pmb{\sigma},r)$ (see \eqref{Eq:Coefficients}) of the expansion coefficient $\mathcal{C}(1,\pmb{\sigma},r)$ is expressed as a product of Lorentz invariants $\epsilon\cdot k$, $\epsilon\cdot\epsilon$ and $k\cdot k$ and constructed by the grapic rule in section \ref{Sec:Expansion}. The gauge invariance states that the amplitude $A(1,2,\dots,r\Vert\,\mathsf{H})$ has to vanish under the replacement $\epsilon_{h}\to k_h$ for any given graviton $h\in\mathsf{H}$. Hence, a relation for pure Yang-Mills amplitudes \cite{Fu:2017uzt} follows
    \bea
    0=\Sl_{\sigma\in\{2,\dots,r-1\}\shuffle\,\text{perms}\,{\mathsf{H}}} \mathcal{C}(1,\pmb{\sigma},r)\Big|_{\epsilon_h\to k_h}A(1,\pmb{\sigma},r).
    \eea
For a given graph in the expansion of $\mathcal{C}(1,\pmb{\sigma},r)$, the graviton $h$ can be either an \textit{internal node} or a \textit{starting point of a chain}. In the former case, the gauge invariance condition is naturally encoded by $F_{h}^{\mu\nu}|_{\epsilon_h\to k_h}=0$, thus this contribution has to vanish. The only nontrivial contributions are those graphs in which the graviton $h$ plays as the starting point of a chain. The gauge invariance condition is then reduced to
    \bea
    0=\Sl_{\sigma\in\{2,\dots,r-1\}\shuffle\,\text{perms}\,{\mathsf{H}}} \Sl_{\mathcal{F}\in{\mathcal{G}^{\pmb{\sigma}}_{\mathsf{H}}[h]}
    }\mathcal{C}^{[\mathcal{F}]}(1,\pmb{\sigma},r)\Big|_{\scriptsize{\epsilon_{h}\to k_{h}}}A(1,\pmb{\sigma},r),\Label{Eq:NewGaugeID0}
    \eea
where $\mathcal{G}^{\pmb{\sigma}}_{\mathsf{H}}[h]$ denotes the set of graphs for permutation $\pmb{\sigma}$, where $h$ plays as starting point of a chain. As shown by examples in \cite{Fu:2017uzt,Du:2017gnh} ({ similar discussions on the gauge invariance relations can be found in \cite{Barreiro:2013dpa,Stieberger:2016lng,Nandan:2016pya,Boels:2016xhc,Chiodaroli:2017ngp,Boels:2017gyc}}), \eqref{Eq:NewGaugeID0} is generated by known BCJ relations, thus it is not new relation beyond known BCJ relations. Nevertheless, a systematical study on the connection between \eqref{Eq:NewGaugeID0} and the standard KK and BCJ relations still deserves future work.

Coefficients in the relation \eqref{Eq:NewGaugeID0} still contain polarizations. To induce a relation where coefficients are only functions of Mandelstam variables $s_{ij}=k_i\cdot k_j$, we should \textit{`turn' all polarizations in the expansion of coefficients to momenta.} One reasonable approach to realize this point is combining gauge invariance conditions with dimensional reduction inspired by CHY formulation. Our discussion is based on the following crucial observations:

\begin{itemize}
\item[(1)] \emph{Gauge invariance conditions for more than one graviton can be imposed simultaneously.} This can be understood from two different aspects. (i) Since the pure Yang-Mills expansion \eqref{Eq:PureYMExpansion} is obtained by applying the recursive expansion \eqref{Eq:RecursiveExpansion} repeatedly, we can take gauge invariance condition for \eqref{Eq:RecursiveExpansion} instead. If we replace $\epsilon_{h_a}$ by $k_{h_a}$ for more than one graviton $h_a\in \mathsf{A}\subseteq\mathsf{H}$ ($\mathsf{A}$ consists of at least two gravitons) on the RHS of \eqref{Eq:RecursiveExpansion}, there is at most one graviton plays as the fiducial one. The polarizations of the rest of the gravitons belonging to $\mathsf{A}$ are contained by either $F^{\mu\nu}$ or an EYM amplitude with fewer gravitons. When replacing  $\epsilon_{h_a}$ by $k_{h_a}$ for all $h_a\in \mathsf{A}$ on the RHS of \eqref{Eq:RecursiveExpansion}, every term has to vanish due to the antisymmetry of  $F^{\mu\nu}$ or/and the gauge invariance condition for EYM amplitudes with fewer gravitons (as an inductive assumption). (ii) In the language of CHY formula \eqref{Eq:CHY}, polarizations are packaged into (reduced) Pffafians.  When the replacement $\epsilon_{h}\to k_{h}$ for a given graviton $h\in \mathsf{H}$ is imposed, the $\Psi_{\mathsf{H}}$ matrix becomes degenerate because two rows/columns coincide with each other (Noting the diagonal entry $C_{h_ah_a}$ for $C$ matrix vanishes due to scattering equation \eqref{Eq:SE}) as shown by the left matrix in the following
    \bea
    \left(
      \begin{array}{ccc|ccc}
        \cdots & \cdots & \cdots & \cdots & \cdots & \cdots \\
        \cdots & {k_{h_a}\cdot k_{h_b}\over z_{h_ah_b}} & \cdots & \cdots & {k_{h_a}\cdot \epsilon_{h_b}\over z_{h_ah_b}} & \cdots \\
        \cdots & \cdots & \cdots & \cdots & \cdots & \cdots \\ \hline
       \cdots & \cdots & \cdots & \cdots & \cdots & \cdots\\
       \cdots & {k_{h_a}\cdot k_{h_b}\over z_{h_ah_b}} & \cdots & \cdots & {k_{h_a}\cdot \epsilon_{h_b}\over z_{h_ah_b}} & \dots \\
         \cdots & \cdots & \cdots & \cdots & \cdots & \cdots\\
      \end{array}
    \right)~~~~~\to~~~~~\left(
      \begin{array}{ccc|ccc}
        \cdots & \cdots & \cdots & \cdots & \cdots & \cdots \\
        \cdots & {k_{h_a}\cdot k_{h_b}\over z_{h_ah_b}} & \cdots & \cdots & {k_{h_a}\cdot k_{h_b}\over z_{h_ah_b}} & \cdots \\
        \cdots & \cdots & \cdots & \cdots & \cdots & \cdots \\ \hline
       \cdots & \cdots & \cdots & \cdots & \cdots & \cdots\\
       \cdots & {k_{h_a}\cdot k_{h_b}\over z_{h_ah_b}} & \cdots & \cdots & {k_{h_a}\cdot k_{h_b}\over z_{h_ah_b}} & \dots \\
         \cdots & \cdots & \cdots & \cdots & \cdots & \cdots\\
      \end{array}
    \right).\Label{Eq:GaugeInvPfaffian}
    \eea
 If we take gauge invariance conditions for more than one graviton, e.g. $h_a$ and $h_b$, the matrix $\Psi$ is also degenerate for the same reason (see the right matrix in \eqref{Eq:GaugeInvPfaffian}), thus the Pfaffian has to vanish.

\item [(2)] \emph{The gauge invariance conditions are independent of dimensions.} This is because the statements (i) and (ii) in (1) hold  for arbitrary dimension space.

\end{itemize}
Having (1) and (2), we can conveniently carry on our discussions in the framework of CHY formula. The recursive and graphic expansions for amplitude reflect corresponding relations for Pfaffians \eqref{Eq:RecPfaffian} and \eqref{Eq:GraphicPfaffian}. Since CHY formula does not depend on the dimension of space, we can extend the Pfaffian  $\Pf\left[\Psi_{\mathsf{H}}\right]$  in the graphic expansion \eqref{Eq:GraphicPfaffian} to $(d+d)$-dimensions by defining $(d+d)$-dimensional polarizations $\mathcal{E}_{h_a}$ (all $h_a\in \mathsf{H}$) and $(d+d)$-dimensional momenta $\mathcal{K}_i$ for all external particles, so that
\bea
\mathcal{E}_{h_a}\cdot \mathcal{K}_{h_a}=0,~~~(\text{for all~} h_a\in \mathsf{H});~~~\mathcal{K}_i\cdot \mathcal{K}_i=0~~~(\text{for all particles $i$});~~~\Sl_{i=1}^{r+s}\mathcal{K}_i=0\Label{Eq:CondtionsdPlusd}
\eea
are satisfied.
According to our observations (1) and (2), the  Pfaffian $\Pf\left[\Psi_{\mathsf{H}}\right]$ in $(d+d)$ dimensions on the LHS of \eqref{Eq:GraphicPfaffian} must vanish under the replacement  $\mathcal{E}_{h_a}\to \mathcal{K}_{h_a}$ for all $h_a\in\mathsf{A}$ where $\mathsf{A}$ is a nonempty subset of $\mathsf{H}$. Consequently, the RHS of the graphic expansion \eqref{Eq:GraphicPfaffian} in $d+d$ dimensions has to vanish when  $\mathcal{E}_{h_a}$ are replaced by $\mathcal{K}_{h_a}$ for all $h_a\in\mathsf{A}\subseteq \mathsf{H}$:
\bea
0=\Sl_{\pmb{\sigma}\in\{2,\dots,r-1\}\shuffle\,\text{perms}\,{\mathsf{H}}} \mathcal{C}(1,\pmb{\sigma},r)\Big|_{\substack{\mathcal{E}_{h_a}\to \mathcal{K}_{h_a}\\\text{for all~}h_a\in\mathsf{A}}}\text{PT}(1,\pmb{\sigma},r).\Label{Eq:GaugeInvdPlusd}
\eea
Once the coefficients $\mathcal{C}(1,\pmb{\sigma},r)$ in the above equation are expressed by graphs (see eq. \eqref{Eq:Coefficients}) and the gauge invariance conditions are imposed, a chain in which any $h_a\in\mathsf{A}\subseteq \mathsf{H}$  plays as an internal node vanishes due to the antisymmetry of the $(d+d)$ dimensional strength tensor $\mathbf{F}_{h_{a}}^{UV}\equiv \mathcal{K}_{h_a}^{U}\mathcal{E}_{h_a}^V-\mathcal{K}_{h_a}^V\mathcal{E}_{h_a}^U$.
Thus only those graphs where all $h_a\in \mathsf{A}$ play as starting points of chains survive. The relation \eqref{Eq:GaugeInvdPlusd} then turns to
    \bea
    0=\Sl_{\pmb{\sigma}\in\{2,\dots,r-1\}\shuffle\,\text{perms}\,{\mathsf{H}}}\biggl[\Sl_{\mathcal{F}\in{\mathcal{G}^{\pmb{\sigma}}_{\mathsf{H}}[\mathsf{A}]}
    }\mathcal{C}^{[\mathcal{F}]}(1,\pmb{\sigma},r)\Big|_{\substack{\scriptsize{\mathcal{E}_{h_a}\to \mathcal{K}_{h_a}}\\\text{for all~}h_a\in\mathsf{A}}}\biggr]\text{PT}(1,\pmb{\sigma},r).\Label{Eq:NewGaugeIDPf}
    \eea
Here, $\mathcal{G}^{\pmb{\sigma}}_{\mathsf{H}}[\mathsf{A}]$ denotes the set of graphs corresponding to the permutation $\pmb{\sigma}$, where all elements in the nonempty subset $\mathsf{A}$ play as starting points of chains (Note that other elements in $\mathsf{H}$ may also be starting points of chains).

The equation \eqref{Eq:NewGaugeIDPf} does not rely on details of $(d+d)$-dimensional polarizations $\mathcal{E}$ and  momenta $\mathcal{K}$, only the conditions \eqref{Eq:CondtionsdPlusd} are required. Thus, we can assign details of polarizations and momenta in $(d+d)$ dimensions appropriately  s.t. \eqref{Eq:CondtionsdPlusd} is satisfied.
A reasonable definition inspired by the dimensional reduction strategy (see \eqref{Eq:CHYreduction}) in the CHY formula is
\bea
    \mathcal{K}_{i}=(k_{i};0),~~~(\text{for all external particles});&~~~~&\mathcal{E}_{h_a}=(0;k_{h_a}),~~~~{h_a\in \mathsf{H}}\Label{Eq:DimensionalReduction}
\eea
which apparently satisfies \eqref{Eq:CondtionsdPlusd}. With this assignment, the coefficients in the gauge invariance condition \eqref{Eq:NewGaugeIDPf} become polynomial functions of Mandelstam variables. When the coefficients $ \mathcal{C}(1,\pmb{\sigma},r)$ in $(d+d)$ dimensions are expressed by the graphic rules and  $\mathcal{E}_{h_a}$ in  $\mathcal{C}(1,\pmb{\sigma},r)$ are replaced by $\mathcal{K}_{h_a}$ ($h_a\in\mathsf{A}\subseteq \mathsf{H}$), chains in the graphs are classified into two types:
 \begin{itemize}
    \item [(i)]\emph{ Type-1} Chains started by  $(d+d)$-dimensional polarizations $\mathcal{E}_{a}$ ($a\in \mathsf{H}\setminus \mathsf{A}$) have the general form
     \bea
     \mathcal{E}_{a}\cdot \mathbf{F}_{h_{i_j}}\cdot \mathbf{F}_{h_{i_{j-1}}}\dots \mathbf{F}_{h_{i_1}}\cdot \mathcal{K}_b.
      \eea
       A chain of this type has to vanish if its length is
    odd, because we cannot avoid a factor of the form $\mathcal{E}_{i}\cdot \mathcal{K}_j$ which is zero in the definition \eqref{Eq:DimensionalReduction}. Thus the length of nonvanishing type-1 chains must be even. When plugging the components \eqref{Eq:DimensionalReduction} into an even-length chain of type-1, we get a chain expressed by $d$-dimensional Mandelstam variables
    \bea
    s_{ah_{i_j}}s_{h_{i_{j}}h_{i_{j-1}}}\dots s_{h_{i_1}b}
    \eea
associated with a factor $(-1)^{j+1\over 2}$, where $j$ is odd. Since the length $L$ of this chain is $j+1$, the prefactor can be given by $(-1)^{L\over 2}$.

  \item [(ii)] \emph{Type-2} Chains started by $(d+d)$-dimensional momenta $\mathcal{K}_a$ have the general form
  \bea
  \mathcal{K}_{a}\cdot \mathbf{F}_{h_{i_j}}\cdot \mathbf{F}_{h_{i_{j-1}}}\dots \mathbf{F}_{h_{i_1}}\cdot \mathcal{K}_b.
  \eea
  A chain of this type vanishes if its length is even, for an even-length type-2 chain must contain  a vanishing factor of the form $\mathcal{E}_{i}\cdot \mathcal{K}_j$.
  Thus the length of nonvanishing type-2 chains are odd. Inserting the  choice of $(d+d)$-dimensional polarizations and momenta \eqref{Eq:DimensionalReduction} into an odd-length chain of this type, we arrive
    \bea
     s_{ah_{i_j}}s_{h_{i_j}h_{i_{j-1}}}\dots s_{h_{i_1}b}
    \eea
    associated with a factor $(-1)^{j\over 2}$, where $j$ is even. The prefactor for this chain is further expressed by the length $L$ of the chain as $(-1)^{L-1\over 2}$.
  \end{itemize}
Collecting all nonzero chains together, we  induce the following  relation for $\text{PT}$ factors in $d$ dimensions from the $(d+d)$-dimensional gauge invariance condition \eqref{Eq:GaugeInvdPlusd}:
 \bea
    0=\Sl_{\sigma\in\{2,\dots,r-1\}\shuffle\,\text{perms}\,{\mathsf{H}}}\Sl_{\mathcal{F}\in{\mathcal{G}'^{\,\pmb{\sigma}}_{\mathsf{H}}[\mathsf{A}]}
    }\mathcal{D}^{[\mathcal{F}]}(1,\pmb{\sigma},r)\text{PT}(1,\pmb{\sigma},r).\Label{Eq:NewGaugeIDPf1}
    \eea
Here, $\mathcal{G}'^{\,\pmb{\sigma}}_{\mathsf{H}}[\mathsf{A}]$ denotes the set of graphs (constructed by the same rule in section \ref{Sec:Expansion}) where \emph{all elements in $\mathsf{A}\subseteq \mathsf{H}$ ($\mathsf{A}\neq\emptyset$) play as starting points of all odd-length chains.} Possible chains of even length must be started by elements in $\mathsf{H}\setminus \mathsf{A}$. For a given permutation $\pmb{\sigma}$ and a given graph $\mathcal{F}\in\mathcal{G}'^{\pmb{\sigma}}_{\mathsf{H}}[\mathsf{A}]$, $\mathcal{D}^{[\mathcal{F}]}(1,\pmb{\sigma},r)$ is obtained by associating chains with factors of the form
\bea
     s_{ah_{i_j}}s_{h_{i_j}h_{i_{j-1}}}\dots s_{h_{i_1}b},
    \eea
in which $a$ and $b$ are the starting points and ending points of a chain, while $h_{i_1}$, ..., $h_{i_j}$ are internal nodes of this chain. Note that the prefactors of all chains in any given graph in $\mathcal{G}'^{\,\pmb{\sigma}}_{\mathsf{H}}[\mathsf{A}]$ together produce a same total factor $(-1)^{{s\over 2}-{1\over 2}{N_{o}}}$, where $s$ is the number of elements in the set $\mathsf{H}$ and equal to the total length of all chains, $N_o$ denotes the number of odd-length chains and is equal to the order of the set $\mathsf{A}$. The total factor thus does not appear in the equation \eqref{Eq:NewGaugeIDPf1}.

To translate the gauge invariance induced relation \eqref{Eq:NewGaugeIDPf1} for Parke-Taylor factors into amplitude relation, we consider the expression
\bea
\Sl_{\pmb{\sigma}\in\{2,\dots,r-1\}\shuffle\,\text{perms}\,{\mathsf{H}}}\Sl_{\mathcal{F}\in{\mathcal{G}'^{\,\pmb{\sigma}}_{\mathsf{H}}[\mathsf{A}]}
    }(-1)^{{(n+1)(n+2)\over 2}}\int d\Omega_{\text{CHY}}\mathcal{D}^{[\mathcal{F}]}(1,\pmb{\sigma},r)\text{PT}(1,\pmb{\sigma},r)\mathcal{I}_R,
\eea
where $\mathcal{I}_R$ can be $\mathcal{I}^{\text{BS}}_R$,  $\mathcal{I}^{\text{YM}}_R$ or $\mathcal{I}^{\text{NLSM}}_R$ in \eqref{Eq:CHYBSintegrand}, \eqref{Eq:CHYYMintegrand} or \eqref{eq:DR} correspondingly. Since the coefficients $\mathcal{D}^{[\mathcal{F}]}(1,\pmb{\sigma},r)$ are independent of the scattering variables, it can be moved outside the integration. The relation for Parke-Taylor factors \eqref{Eq:NewGaugeIDPf1} then gives the following gauge invariance induced amplitude relations
 \bea
    \boxed{0=\Sl_{\pmb{\sigma}\in\{2,\dots,r-1\}\shuffle\,\text{perms}\,{\mathsf{H}}}\Sl_{\mathcal{F}\in{\mathcal{G}'^{\,\pmb{\sigma}}_{\mathsf{H}}[\mathsf{A}]}
    }\mathcal{D}^{[\mathcal{F}]}(1,\pmb{\sigma},r)A(1,\pmb{\sigma},r)\Label{Eq:NewGaugeIDAmp1}},
    \eea
for any nonempty $\mathsf{A}$ ($\mathsf{A}\subseteq \mathsf{H}$).
\subsection{Examples for the gauge invariance induced relation \eqref{Eq:NewGaugeIDAmp1}}

Now let us present several examples of the gauge invariance induced amplitude relation \eqref{Eq:NewGaugeIDAmp1}.
\begin{figure}
\centering
\includegraphics[width=7in]{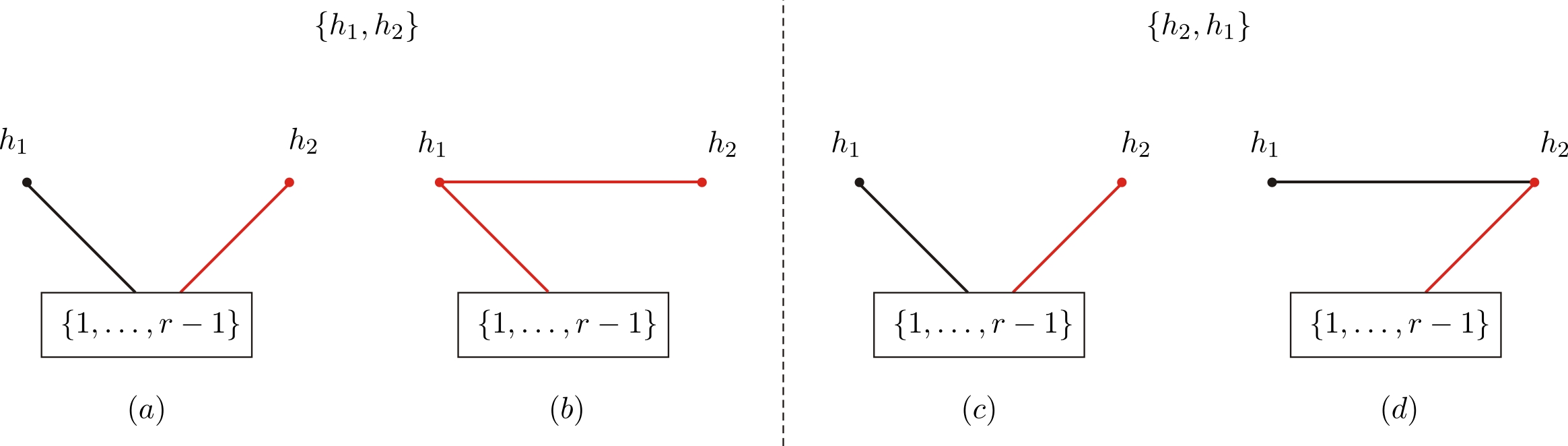}
\caption{All possible graphs with $\mathsf{H}=\{h_1,h_2\}$ and reference order $\mathsf{R}=\{h_1,h_2\}$. Graphs $(a)$ and $(b)$ correspond to the permutations $\{2,\dots,r-1\}\shuffle\{h_1,h_2\}$, while graphs $(a)$ and $(b)$ correspond to the permutations $\{2,\dots,r-1\}\shuffle\{h_2,h_1\}$.}\label{Fig:Figure1}
\end{figure}
\subsubsection{$\mathsf{H}=\{h_1,h_2\}$}

The first example is given by $\mathsf{H}=\{h_1,h_2\}$. If the reference order is fixed as $\mathsf{R}=\{h_1,h_2\}$, all graphs given by the graphic rule in section \ref{Sec:Expansion} are displayed in figure \ref{Fig:Figure1}. The graphs $(a)$, $(b)$ in figure \ref{Fig:Figure1} contribute to permutations $\{2,\dots,r-1\}\shuffle\{h_1,h_2\}$, while $(c)$, $(d)$ contribute to the relative order $\{2,\dots,r-1\}\shuffle\{h_2,h_1\}$.

In the gauge invariance induced relation \eqref{Eq:NewGaugeIDAmp1}, the nonempty subset $\mathsf{A}$  cannot contain only one element because the total length of all chains is an even number $2$. If $\mathsf{A}$ contains for example $h_1$, {\it i.e.}, there is an odd-length chain started by $h_1$, we must have another odd-length chain started by $h_2$ so that the total length of all chains is even. Thus the nonempty subset $\mathsf{A}$ of $\mathsf{H}$ can only be chosen as $\{h_1,h_2\}$ while $h_1$ and $h_2$ are starting points of two length-1 chains in this example. The graph $(b)$ which contains a length-2 chain does not appear in our gauge invariance induced relation. The relation \eqref{Eq:NewGaugeIDAmp1} for $\mathsf{A}=\{h_1,h_2\}$ reads
\bea
0&=&\Sl_{\pmb{\sigma}}s_{h_2X_{h_2}}s_{h_1X_{h_1}}A(1,\pmb{\sigma}\in\{2,\dots,r-1\}\shuffle\{h_1,h_2\},r)\nn
&&+\Sl_{\pmb{\sigma}}s_{h_2X_{h_2}}(s_{h_1X_{h_1}}+s_{h_1h_2})A(1,\pmb{\sigma}\in\{2,\dots,r-1\}\shuffle\{h_2,h_1\},r),\Label{Eq:GaugeInducedExample1}
\eea
where $s_{aX_a}\equiv\Sl_{\scriptsize\substack{i\in\{1,2,\dots,r-1\}\\\text{s.t.} \sigma^{-1}(i)<\sigma^{-1}(a)}}s_{ai}$.
This relation is in agreement with a fundamental BCJ relation.

\subsubsection{$\mathsf{H}=\{h_1,h_2,h_3\}$}

We consider the examples with $\mathsf{H}=\{h_1,h_2,h_3\}$. For the reference order $\mathsf{R}=\{h_1,h_2,h_3\}$, all possible graphs constructed by the graphic rules are provided by figure \ref{Fig:5ptGraphs} in appendix \ref{App:5Pt}. For any graph, the total length of all chains must be $3$. As a result, the nonempty subset $\mathsf{A}$ in the relation \eqref{Eq:NewGaugeIDAmp1} can only contain odd number of elements, {\it i.e.},  $\mathsf{A}$ can be $\{h_1\}$, $\{h_2\}$, $\{h_3\}$ or $\{h_1,h_2,h_3\}$.

\subsubsection*{$\mathsf{A}=\{h_1\}$}
 If $\mathsf{A}$ contains only one element $h_1$. Then $h_1$ must leads to a length-1 chain while $h_3$ must leads to a length-2 chain $s_{h_3h_2}s_{h_2a}$ with an internal node $h_2$. Among the graphs in figure \ref{Fig:5ptGraphs}, only $(a5)$  (for the relative order $\{h_1,h_2,h_3\}$), $(c3)$, $(c4)$ (for the relative order $\{h_2,h_1,h_3\}$) and $(d2)$, $(d4)$, $(d6)$ (for the relative order $\{h_2,h_3,h_1\}$)  contribute. Hence the relation for $\mathsf{A}=\{h_1\}$ is
\bea
0&=&\Sl_{\pmb{\sigma}}s_{h_1X_{h_1}}s_{h_3h_2}s_{h_2X_{h_2}}A(1,\pmb{\sigma}\in\{2,\dots,r-1\}\shuffle\{h_1,h_2,h_3\},r)\nn
&+&\Sl_{\pmb{\sigma}}(s_{h_1X_{h_1}}+s_{h_1h_2})s_{h_3h_2}s_{h_2X_{h_2}}A(1,\pmb{\sigma}\in\{2,\dots,r-1\}\shuffle\{h_2,h_1,h_3\},r)\nn
&+&\Sl_{\pmb{\sigma}}(s_{h_1X_{h_1}}+s_{h_1h_2}+s_{h_1h_3})s_{h_3h_2}s_{h_2X_{h_2}}A(1,\pmb{\sigma}\in\{2,\dots,r-1\}\shuffle\{h_2,h_3,h_1\},r).\Label{Eq:GaugeInducedExample2}
\eea
This relation is consistent with a fundamental BCJ relation.

\subsubsection*{$\mathsf{A}=\{h_2\}$}
If $\mathsf{A}=\{h_2\}$, $h_2$ must be the starting point of a length-1 chain under the choice of reference order $\mathsf{R}=\{h_1,h_2,h_3\}$, while $h_3$ must start a length-2 chain with the internal node $h_1$. The graphs $(a3)$, $(a4)$, $(b2)$, $(b4)$, $(b6)$ and $(c5)$ have nonvanishing contributions and the relation \eqref{Eq:NewGaugeIDAmp1} gives
\bea
0&=&\Sl_{\pmb{\sigma}}s_{h_2X_{h_2}} s_{h_3h_1}s_{h_1X_{h_1}}A(1,\pmb{\sigma}\in\{2,\dots,r-1\}\shuffle\{h_2,h_1,h_3\},r)\nn
&+&\Sl_{\pmb{\sigma}}(s_{h_2X_{h_2}}+s_{h_2h_1})s_{h_3h_1}s_{h_1X_{h_1}}A(1,\pmb{\sigma}\in\{2,\dots,r-1\}\shuffle\{h_1,h_2,h_3\},r)\nn
&+&\Sl_{\pmb{\sigma}}(s_{h_2X_{h_2}}+s_{h_2h_1}+s_{h_2h_3})s_{h_3h_1}s_{h_1X_{h_1}}A(1,\pmb{\sigma}\in\{2,\dots,r-1\}\shuffle\{h_1,h_3,h_2\},r).\Label{Eq:GaugeInducedExample3}
\eea
Again, the vanish of RHS can be considered as a result of fundamental BCJ relation.
\subsubsection*{$\mathsf{A}=\{h_3\}$}
If $\mathsf{A}=\{h_3\}$, the element $h_3$ can start either a length-$3$ chain or a length-$1$ chain. In the former case, both $h_1$ and $h_2$ must be internal nodes of the length-3 chain ($(a6)$ and $(c6)$ in figure \ref{Fig:5ptGraphs}), while in the latter case $h_2$ must start a length-$2$ chain with $h_1$ as the internal node ((a2), (b3), (e5) and (e6) in figure \ref{Fig:5ptGraphs}). All together, the relation  \eqref{Eq:NewGaugeIDAmp1} turns to
\bea
0&=&\Sl_{\pmb{\sigma}}(s_{h_3h_2}s_{h_2h_1}s_{h_1X_{h_1}}+s_{h_3X_{h_3}}s_{h_2h_1}s_{h_1X_{h_1}})A(1,\pmb{\sigma}\in\{2,\dots,r-1\}\shuffle\{h_1,h_2,h_3\},r)\nn
&&+\Sl_{\pmb{\sigma}}s_{h_3h_1}s_{h_1h_2}s_{h_2X_{h_2}}A(1,\pmb{\sigma}\in\{2,\dots,r-1\}\shuffle\{h_2,h_1,h_3\},r)\nn
&&+\Sl_{\pmb{\sigma}}s_{h_3X_{h_3}}s_{h_2h_1}s_{h_1X_{h_1}}A(1,\pmb{\sigma}\in\{2,\dots,r-1\}\shuffle\{h_1,h_3,h_2\},r)\nn
&&+\Sl_{\pmb{\sigma}}s_{h_3X_{h_3}}s_{h_2h_1}(s_{h_1X_{h_1}}+s_{h_1h_3})A(1,\pmb{\sigma}\in\{2,\dots,r-1\}\shuffle\{h_3,h_1,h_2\},r),\Label{Eq:GaugeInducedExample4}
\eea
which is not as trivial as previous examples. One can check this identity by expanding all amplitudes in terms of BCJ basis amplitudes.

\subsubsection*{$\mathsf{A}=\{h_1,h_2,h_3\}$}
Now we consider the case $\mathsf{A}=\{h_1,h_2,h_3\}$, for which all elements in $\mathsf{H}$ play as starting points of odd-length chains. The only possibility is that all chains are of length $1$. The relation \eqref{Eq:NewGaugeIDAmp1} then gives rise
\bea
0&=&\Sl_{\pmb{\sigma}}s_{h_1X_{h_1}}s_{h_2X_{h_2}}s_{h_3X_{h_3}}A(1,\pmb{\sigma}\in\{2,\dots,r-1\}\shuffle\{h_1,h_2,h_3\},r)\nn
&&+\Sl_{\pmb{\sigma}}(s_{h_1X_{h_1}}+s_{h_1h_2})s_{h_2X_{h_2}}s_{h_3X_{h_3}}A(1,\pmb{\sigma}\in\{2,\dots,r-1\}\shuffle\{h_2,h_1,h_3\},r)\nn
&&+\Sl_{\pmb{\sigma}}(s_{h_1X_{h_1}}+s_{h_1h_2}+s_{h_1h_3})s_{h_2X_{h_2}}s_{h_3X_{h_3}}A(1,\pmb{\sigma}\in\{2,\dots,r-1\}\shuffle\{h_2,h_3,h_1\},r)\nn
&&+\Sl_{\pmb{\sigma}}s_{h_1X_{h_1}}(s_{h_2X_{h_2}}+s_{h_2h_3})s_{h_3X_{h_3}}A(1,\pmb{\sigma}\in\{2,\dots,r-1\}\shuffle\{h_1,h_3,h_2\},r)\nn
&&+\Sl_{\pmb{\sigma}}(s_{h_1X_{h_1}}+s_{h_1h_3})(s_{h_2X_{h_2}}+s_{h_2h_3})s_{h_3X_{h_3}}A(1,\pmb{\sigma}\in\{2,\dots,r-1\}\shuffle\{h_3,h_1,h_2\},r)\nn
&&+\Sl_{\pmb{\sigma}}(s_{h_1X_{h_1}}+s_{h_1h_3}+s_{h_1h_2})(s_{h_2X_{h_2}}+s_{h_2h_3})s_{h_3X_{h_3}}A(1,\pmb{\sigma}\in\{2,\dots,r-1\}\shuffle\{h_3,h_2,h_1\},r).\Label{Eq:GaugeInducedExample5}
\eea
The RHS of the above relation gets contributions from eighteen graphs $(a1)$, $(b1)$, $(b5)$, $(c1)$, $(c2)$, $(d1)$, $(d3)$, $(d5)$,  $(e1)$, $(e2)$, $(e3)$, $(e4)$, $(f1)$, $(f2)$, $(f3)$, $(f4)$, $(f5)$ and $(f6)$. Both  the sum of the first three rows and  the sum of the last three rows vanish due to fundamental BCJ relation.

\subsubsection{$\mathsf{H}=\{h_1,h_2,h_3,h_4\}$}

We consider a much more nontrivial case with $\mathsf{H}=\{h_1,h_2,h_3,h_4\}$ as the last example. The nonempty subset in \eqref{Eq:NewGaugeIDAmp1} is chosen as $\mathsf{A}=\{h_3,h_4\}$ and the reference order is chosen as $\mathsf{R}=\{h_1,h_2,h_3,h_4\}$. If $h_4$ ($h_3$) is starting point of a length-$3$ chain, $h_3$ ($h_4$) must be starting point of a length-$1$ chain. Such graphs contain only two chains; If both $h_4$ and $h_3$ are starting points of length-$1$ chains, we must also have an length-$2$ chain of the form $s_{h_2h_1}s_{h_1Y_{h_1}}$. The coefficients for all possible permutations are displayed as follows ($\{h_1h_2h_3h_4\}$ is used to denote the permutation $1,\{2,\dots,r-1\}\shuffle\{h_1,h_2,h_3,h_4\},r$ for short)
\bea
&&\{h_3h_1h_2h_4\}:s_{h_4h_2}s_{h_2h_1}s_{h_1X_{h_1}}s_{h_3X_{h_3}}+s_{h_4X_{h_4}}s_{h_3X_{h_3}}s_{h_2h_1}(s_{h_1X_{h_1}}+s_{h_1h_3}),\nn
&&\{h_1h_3h_2h_4\}:s_{h_4h_2}s_{h_2h_1}s_{h_1X_{h_1}}(s_{h_3X_{h_3}}+s_{h_3h_1})+s_{h_4X_{h_4}}s_{h_3X_{h_3}}s_{h_2h_1}s_{h_1X_{h_1}},\nn
&&\{h_1h_2h_3h_4\}: s_{h_4h_2}s_{h_2h_1}s_{h_1X_{h_1}}(s_{h_3X_{h_3}}+s_{h_3h_1}+s_{h_3h_2})+s_{h_4X_{h_4}}(s_{h_3X_{h_3}}+s_{h_3h_2})s_{h_2h_1}s_{h_1X_{h_1}},\nn
&&\{h_1h_2h_4h_3\}:s_{h_4h_2}s_{h_2h_1}s_{h_1X_{h_1}}(s_{h_3X_{h_3}}+s_{h_3h_1}+s_{h_3h_2}+s_{h_3h_4})+s_{h_4X_{h_4}}(s_{h_3X_{h_3}}+s_{h_3h_2}+s_{h_3h_4})s_{h_2h_1}s_{h_1X_{h_1}},\nn
&&\{h_3h_1h_4h_2\}:s_{h_4X_{h_4}}s_{h_3X_{h_3}}s_{h_2h_1}(s_{h_1X_{h_1}}+s_{h_1h_3}),~~~~\{h_1h_3h_4h_2\}:s_{h_4X_{h_4}}s_{h_3X_{h_3}}s_{h_2h_1}s_{h_1X_{h_1}},\nn
&&\{h_1h_4h_3h_2\}:s_{h_4X_{h_4}}(s_{h_3X_{h_3}}+s_{h_3h_4})s_{h_2h_1}s_{h_1X_{h_1}},~~~~\{h_1h_4h_2h_3\}:s_{h_4X_{h_4}}(s_{h_3X_{h_3}}+s_{h_3h_4}+s_{h_3h_2})s_{h_2h_1}s_{h_1X_{h_1}},\nn
&&\{h_3h_4h_1h_2\}:s_{h_4X_{h_4}}s_{h_3X_{h_3}}s_{h_2h_1}(s_{h_1X_{h_1}}+s_{h_1h_3}+s_{h_1h_4}),\nn
&&\{h_4h_3h_1h_2\}:s_{h_4X_{h_4}}(s_{h_3X_{h_3}}+s_{h_3h_4})s_{h_2h_1}(s_{h_1X_{h_1}}+s_{h_1h_3}+s_{h_1h_4}),\nn
&&\{h_4h_1h_3h_2\}:s_{h_4X_{h_4}}(s_{h_3X_{h_3}}+s_{h_3h_4})s_{h_2h_1}(s_{h_1X_{h_1}}+s_{h_1h_4}),\nn
&&\{h_4h_1h_2h_3\}:s_{h_4X_{h_4}}(s_{h_3X_{h_3}}+s_{h_3h_4}+s_{h_3h_2})s_{h_2h_1}s_{h_1X_{h_1}},\nn
&&\{h_3h_2h_1h_4\}:s_{h_4h_1}s_{h_1h_2}s_{h_2X_{h_2}}s_{h_3X_{h_3}},~~~~~~~~~~~~~~~~~~\{h_2h_3h_1h_4\}:s_{h_4h_1}s_{h_1h_2}s_{h_2X_{h_2}}(s_{h_3X_{h_3}}+s_{h_3h_2}),\nn
&&\{h_2h_1h_3h_4\}:s_{h_4h_1}s_{h_1h_2}s_{h_2X_{h_2}}(s_{h_3X_{h_3}}+s_{h_3h_2}+s_{h_3h_1})+s_{h_4X_{h_4}}s_{h_3h_1}s_{h_1h_2}s_{h_2X_{h_2}},\nn
&&\{h_2h_1h_4h_3\}:s_{h_4h_1}s_{h_1h_2}s_{h_2X_{h_2}}(s_{h_3X_{h_3}}+s_{h_3h_2}+s_{h_3h_1}+s_{h_3h_4})+s_{h_4X_{h_4}}s_{h_3h_1}s_{h_1h_2}s_{h_2X_{h_2}}\nn
&&\{h_2h_4h_1h_3\}:s_{h_4X_{h_4}}s_{h_3h_1}s_{h_1h_2}s_{h_2X_{h_2}},~~~~~~~~~~~~~~~~~~\{h_4h_2h_1h_3\}:s_{h_4X_{h_4}}s_{h_3h_1}s_{h_1h_2}(s_{h_2X_{h_2}}+s_{h_2h_4}).\Label{Eq:GaugeInducedExample6}
\eea

\subsection{The boundary case $\mathsf{A}=\mathsf{H}$ and partial momentum kernel}
When we set $\mathsf{A}=\mathsf{H}$, every graph in the gauge invariance induced relation \eqref{Eq:NewGaugeIDAmp1} only contains length-1 chains (as shown by examples \eqref{Eq:GaugeInducedExample1} and \eqref{Eq:GaugeInducedExample5}).
Then the relation \eqref{Eq:NewGaugeIDAmp1} becomes
\bea
0=\Sl_{\pmb{\sigma}\in\{2,\dots,r-1\}\shuffle\,\text{perms}\,{\mathsf{H}}}\Sl_{\mathcal{F}\in{\mathcal{G}'^{\pmb{\sigma}}_{\mathsf{H}}[\mathsf{H}]}
    }\mathcal{D}^{[\mathcal{F}]}(1,\pmb{\sigma},r)A(1,\pmb{\sigma},r). \Label{Eq:NewGaugeIDAmpBoundary}
\eea
Assuming that the reference order is $\mathsf{R}=\left\{h_{\rho(1)},h_{\rho(2)},\dots,h_{\rho(s)}\right\}$, let us analyze the coefficients in the above equation in more detail.
A length-1 chain started by $h_{\rho(s)}$ can end at any gluon $l_s\in\{1,\dots,r-1\}$ s.t. $\sigma^{-1}(l_s)<\sigma^{-1}(h_{\rho(s)})$ and is associated with a factor $s_{h_{\rho(s)}l_s}$.
A length-1 chain started by $h_{\rho(s-1)}$ can end at any element $l_{s-1}\in \{1,\dots,r-1\}\cup\{h_{\rho(s)}\}$ s.t., $\sigma^{-1}(l_{s-1})<\sigma^{-1}(h_{\rho(s-1)})$ and is associated with a factor $s_{h_{\rho(s-1)}l_{s-1}}$. This observation can be extended to arbitrary case: a length-1 chain started by $h_{\rho(i)}$ in \eqref{Eq:NewGaugeIDAmpBoundary} can end at any $l_{i}\in \{1,\dots,r-1\}\cup\{h_{\rho(i+1)},\dots,h_{\rho(s)}\}$ s.t.,
$\sigma^{-1}(l_{i})<\sigma^{-1}(h_{\rho(i)})$.
The coefficient for given permutation $\pmb{\sigma}$ then reads
\bea
\Sl_{\mathcal{F}\in{\mathcal{G}'^{\pmb{\sigma}}_{\mathsf{H}}[\mathsf{H}]}
    }\mathcal{D}^{[\mathcal{F}]}(1,\pmb{\sigma},r)&=&\Sl_{\substack{l_i\in\{1,2,\dots,r-1\}\cup\{h_{\rho(i+1)},\dots,h_{\rho(s)}\}\\\text{s.t.\,} \sigma^{-1}(l_i)<\sigma^{-1}(h_{\rho(i)})\,\text{for all}\,i=1,\dots,s}}s_{h_{\rho(1)}l_1}s_{h_{\rho(2)}l_2}\dots s_{h_{\rho(s)}l_s}.\Label{Eq:BoundaryCoefficient}
\eea

An interesting observation  is that we can reexpress the coefficient \eqref{Eq:BoundaryCoefficient} by defining `partial momentum kernel'. Given two permutations $\pmb{\sigma}$ and $\pmb{\rho}$ of elements in $\{2,\dots,m\}$  and a nonempty subset  $\mathsf{H}$ of $\{2,\dots,m\}$, the partial momentum kernel $\W S_{\mathsf{H}}[\pmb{\sigma}|\pmb{\rho}]$ is defined by
\bea
\W S_{\mathsf{H}}[\pmb{\sigma}|\pmb{\rho}]\equiv\prod\limits_{a\in \mathsf{H}}\biggl[s_{a1}+\Sl_{l\in \{2,\dots,m\}}\theta(\sigma^{-1}(a)-\sigma^{-1}(l))\theta(\rho^{-1}(a)-\rho^{-1}(l))s_{al}\biggr],\Label{Eq:PartialMomentumKernal}
\eea
where $\sigma^{-1}(a)$ and $\rho^{-1}(a)$ denote the positions of $a$ in the permutations $\pmb{\sigma}$ and $\pmb{\rho}$ respectively .
Given $a\in \mathsf{H}$ and $l\in \{2,\dots,m\}$, the product of two step functions in \eqref{Eq:PartialMomentumKernal} is $1$ if both $\sigma^{-1}(a)>\sigma^{-1}(l)$ and $\rho^{-1}(a)>\rho^{-1}(l)$ are satisfied, otherwise $0$.
Explicit examples of the partial momentum kernel are given as
\bea
\W S_{\{2\}}[2345|2543]&=&s_{21},~~~~~~~~~~~~~~~~~~~~~~~~~~~~~~~~~\,\W S_{\{3\}}[2345|5423]=s_{31}+s_{32}, \nn
\W S_{\{2,5\}}[2345|4235]&=&s_{21}(s_{51}+s_{52}+s_{53}+s_{54}),~\W S_{\{2,3,4\}}[2345|3542]=s_{21}s_{31}(s_{41}+s_{43}).
\eea
There are many useful properties satisfied by partial momentum kernels:
\begin{itemize}
\item [(i)] Partial momentum kernel $\W S_{\mathsf{H}}[\pmb{\sigma}|\pmb{\rho}]$ is symmetric under exchanging of permutations $\pmb{\sigma}$ and $\pmb{\rho}$, {\it i.e.},
\bea
\W S_{\mathsf{H}}[\pmb{\sigma}|\pmb{\rho}]=\W S_{\mathsf{H}}[\pmb{\rho}|\pmb{\sigma}].\Label{Eq:prop1}
\eea
\item [(ii)] If the subset $\mathsf{H}$ is chosen as the full set $\{2,\dots,m\}$, we arrive the usual momentum kernel
\bea
\W S_{\{2,\dots,m\}}[\pmb{\sigma}_{2,m}|\pmb{\rho}_{2,m}]=S[\pmb{\sigma}_{2,m}|\pmb{\rho}_{2,m}].\Label{Eq:prop2}
\eea
\item [(iii)] Assuming that $\pmb{\rho}_{\mathsf{B}}$ and $\pmb{\rho}'_{\mathsf{B}}$ are two permutations of elements of a set $\mathsf{B}$, while $\pmb{\rho}_{\mathsf{C}}$ is a permutation of elements of $\mathsf{C}$, the partial momentum kernel $\W S_{\mathsf{C}}\left[\pmb{\rho}_{\mathsf{B}},\pmb{\rho}_{\mathsf{C}}|\pmb{\sigma}_{\mathsf{B}}\shuffle\pmb{\sigma}_{\mathsf{C}}\right]$ satisfies
\bea
\W S_{\mathsf{C}}\left[\pmb{\rho}_{\mathsf{B}},\pmb{\rho}_{\mathsf{C}}|\pmb{\sigma}_{\mathsf{B}}\shuffle\pmb{\sigma}_{\mathsf{C}}\right]=\W S_{\mathsf{C}}\left[\pmb{\rho}'_{\mathsf{B}},\pmb{\rho}_{\mathsf{C}}|\pmb{\sigma}_{\mathsf{B}}\shuffle\pmb{\sigma}_{\mathsf{C}}\right].\Label{Eq:prop3}
\eea

\item [(iv)] The following property which relates usual momentum kernel and partial momentum kernel will be useful in the coming sections:
\bea
S\left[\pmb{\rho}_{\mathsf{B}},\pmb{\rho}_{\mathsf{C}}|\pmb{\sigma}_{\mathsf{B}}\shuffle\pmb{\sigma}_{\mathsf{C}}\right]=S\left[\pmb{\rho}_{\mathsf{B}}|\pmb{\sigma}_{\mathsf{B}}\right]\W S_{\mathsf{C}}[\pmb{\rho}_{\mathsf{B}},\pmb{\rho}_{\mathsf{C}}|\pmb{\sigma}_{\mathsf{B}}\shuffle\pmb{\sigma}_{\mathsf{C}}].\Label{Eq:prop4}
\eea
\end{itemize}

Having defined the partial momentum kernel \eqref{Eq:PartialMomentumKernal} and choosing the the reference order as $\mathsf{R}=\{h_{\rho(1)},h_{\rho(2)},\dots,h_{\rho(s)}\}$, we naturally write the coefficient \eqref{Eq:BoundaryCoefficient} as
\bea
    \Sl_{\mathcal{F}\in{\mathcal{G}'^{\pmb{\sigma}}_{\mathsf{H}}[\mathsf{H}]}
    }\mathcal{D}^{[\mathcal{F}]}(1,\pmb{\sigma},r)=\W S_{\mathsf{H}}[\pmb{\sigma}|2,\dots,r-1,h_{\rho(s)},h_{\rho(s-1)},\dots,h_{\rho(1)}].
    \eea
The relation  \eqref{Eq:NewGaugeIDAmpBoundary} for $\mathsf{A}=\mathsf{H}$ is then conveniently given by
\bea
    0=\Sl_{\pmb{\sigma}\in\{2,\dots,r-1\}\shuffle\text{perms~} \mathsf{H}}\W S_{\mathsf{H}}[\pmb{\sigma}|2,\dots,r-1,h_{\rho(s)},h_{\rho(s-1)},\dots,h_{\rho(1)}]A(1,\pmb{\sigma},r).\Label{Eq:NewGaugeIDAmpBoundary1}
    \eea
For the cases with $\mathsf{H}=\{h_1,h_2\}$ and $\mathsf{H}=\{h_1,h_2,h_3\}$, \eqref{Eq:NewGaugeIDAmpBoundary1} returns to the examples \eqref{Eq:GaugeInducedExample1} and \eqref{Eq:GaugeInducedExample5} respectively. In fact, the relation \eqref{Eq:NewGaugeIDAmpBoundary1} is consistent with the following fundamental BCJ relation for given permutation $\pmb{\eta}\in\{2,\dots,r-1\}\shuffle\text{perms\,}\{\mathsf{H}\setminus \{h_{\rho(s)}\}\}$
\bea
0&=&s_{h_{\rho(s)}1}A(1,h_{\rho(s)},\eta(1),\eta(2),\dots,\eta(r+s-2),r)\nn
&&+(s_{h_{\rho(s)}1}+s_{h_{\rho(s)}\eta(1)})A(1,\eta(1),h_{\rho(s)},\eta(2),\dots,\eta(r+s-2),r)\nn
&&+\dots+(s_{h_{\rho(s)}1}+s_{h_{\rho(s)}\eta(1)}+\dots+s_{h_{\rho(s)}\eta(r+s-2)})A(1,\eta(1),\eta(2),\dots,\eta(r+s-2),h_{\rho(s)},r).
\eea

\section{Three types of BCJ numerators in NLSM}\label{Sec:NumeratorsNLSM1}
As an application of the gauge invariance induced relation \eqref{Eq:NewGaugeIDAmp1}, we will prove the equivalence between distinct approaches to scattering amplitudes in NLSM: (i) traditional Feynman diagrams, (ii) the CHY formula and (iii) the Abelian Z theory in the remaining sections.  The starting point of our proof is the fact that all three approaches result \emph{dual DDM formula}
\bea
M(1,\dots,n)=\Sl_{\pmb{\sigma}\in S_{n-2}}n_{1|\pmb{\sigma}|n} A(1,\pmb{\sigma},n),~~~(\text{$n$ is even})\Label{Eq:DualDDM}
\eea
with distinct (DF, CMS and DT) expressions of BCJ numerators $n_{1|\pmb{\sigma}|n}$ (as polynomial functions of Mandelstam variables). The  $A(1,\pmb{\sigma},n)$ in \eqref{Eq:DualDDM} are bi-scalar amplitudes. Thus the three approaches are equivalent to each other if and only if the following relations for bi-scalar amplitudes are satisfied:
\bea
\Sl_{\pmb{\sigma}\in S_{n-2}}n^{\text{DF}}_{1|\pmb{\sigma}|n}A(1,\pmb{\sigma},n)=\Sl_{\pmb{\sigma}\in S_{n-2}}n^{\text{CMS}}_{1|\pmb{\sigma}|n} A(1,\pmb{\sigma},n)=\Sl_{\pmb{\sigma}\in S_{n-2}}n^{\text{DT}}_{1|\pmb{\sigma}|n}A(1,\pmb{\sigma},n).\Label{Eq:Equivalence}
\eea
 We will review the three types of BCJ numerators in this section and  prove the equivalence condition \eqref{Eq:Equivalence} by using \eqref{Eq:NewGaugeIDAmp1} in sections \ref{Sec:DTCMS} and \ref{Sec:DFCMS}.

\subsection{Three distinct constructions of BCJ numerators in NLSM }\label{Sec:NumeratorsNLSM}
Now let us review the DF, CMS and DT types of BCJ numerators which correspond  to the Feynman diagram approach, Abelian Z theory and CHY formula.

\subsubsection*{The DF type numerators}
The DF type  BCJ numerator was derived by applying off-shell extended BCJ relation \cite{Chen:2013fya, Du:2016tbc}, which is based on Berends-Giele recursion (thus Feynman diagrams).
The explicit expression of DF type BCJ numerator is given by a proper combination of momentum kernel:\footnote{We adjust the total sign by $(-1)$ to agree with the CMS type numerators.}
\bea
n^{\text{DF}}_{1|\pmb{\sigma}|n}=(-1)\Sl_{\pmb{\rho}\in \mathsf{\Gamma}}S[\pmb{\sigma}|\pmb{\rho}],\Label{Eq:DFform}
\eea
where we summed over permutations $\pmb{\rho}$ in  $\mathsf{\Gamma}$ which is defined as the collection of permutations satisfying the following conditions. For any $a\in\{2,\dots,n-1\}$, we assume $b$ ($c$) is the nearest element on the LHS (RHS) of $a$ in the permutation $\pmb{\rho}$, which satisfies $\sigma^{-1}(b)>\sigma^{-1}(a)$ ($\sigma^{-1}(c)>\sigma^{-1}(a)$) \footnote{
Here $1$ and $n$ are correspondingly considered as the first and the last elements in both permutations $\pmb{\sigma}$ and $\pmb{\rho}$.
There is always a particle  $n$ (maybe not the nearest ) on the RHS and  LHS of $a$ in the permutation $\pmb{\rho}$  s.t.  $\sigma^{-1}(n)=n>\sigma^{-1}(a)$ in the sense of cyclicity, see  \cite{ Du:2016tbc}.} . The permutations $\pmb{\rho}$ in the DF type numerator \eqref{Eq:DFform} are those satisfying either of the following two conditions: (i) There are odd number of elements between  $a$, $b$ as well as $a$, $c$ in the permutation $\pmb{\rho}$. (ii) There is no element between both $a$, $b$ and $a$, $c$ in the permutation $\pmb{\rho}$. 
Explicit examples are given as
\bea
n^{\text{DF}}_{1|23|4}&=&S[23|32]=-s_{21}s_{31},\Label{Eq:DF4Pt}\\
n^{\text{DF}}_{1|2345|6}&=&-(S[2345|5243]+S[2345|5342]+S[2345|4352]+S[2345|4253]+S[2345|3254])\Label{Eq:DF6Pt}\nn
&=&(-1)\Bigl[s_{51}\left(s_{41}+s_{42}\right)(s_{31}+s_{32})s_{21}+s_{51}\left(s_{41}+s_{43}\right)s_{31}s_{21}\nn
&&+(s_{51}+s_{54}+s_{53})s_{41}s_{31}s_{21}+(s_{51}+s_{54}+s_{52})s_{41}(s_{31}+s_{32})s_{21}\nn
&&+(s_{51}+s_{52}+s_{53})(s_{41}+s_{42}+s_{43})s_{31}s_{21}\Bigr].
\eea

\subsubsection*{The CMS type numerators}
The CMS type BCJ numerator, which comes from Abelian Z theory \cite{Carrasco:2016ldy}, expresses each numerator in dual DDM decomposition by only one momentum kernel:
\bea
n^{\text{CMS}}_{1|\pmb{\sigma}|n}=(-1)^{n\over 2}S[\sigma(2),\sigma(3),\dots,\sigma(n-1)|\sigma(2),\sigma(3),\dots,\sigma(n-1)].\Label{Eq:CMSform}
\eea
 Explicit expressions for four- and six-point cases are
\bea
n^{\text{CMS}}_{1|23|4}&=&S[23|23]=s_{21}(s_{31}+s_{32})\nn
n^{\text{CMS}}_{1|2345|6}&=&S[2345|2345]=s_{21}(s_{31}+s_{32})(s_{41}+s_{42}+s_{43})(s_{51}+s_{52}+s_{53}+s_{54}).
\eea
It is worthy emphasizing that both DF  and CMS types BCJ numerators manifest the relabeling symmetry of $n-2$ elements, {\it i.e.}, $n_{1|\sigma(2),\dots,\sigma(n-1)|n}$ can be obtained from $n_{1|2,\dots,n-1|n}$ by the replacement $2,3,\dots,n-1\to\sigma(2),\sigma(3),\dots,\sigma(n-1)$.

\subsubsection*{The DT type numerators}
Being different from the previous two constructions, the DT type numerator which is based on the graphic expansion of amplitudes and the dimensional reduction in CHY formula is not a symmetric form. This type of BCJ numerators are expanded by graphic rule instead of momentum kernels. The construction of $n^{\text{DT}}_{1|\pmb{\sigma}|n}$ is given by
\begin{itemize}
\item Consider $1$ as the root of a tree and define a reference order of elements in $\{2,\dots,n-1\}$, say $\mathsf{R}\equiv\{\rho(1),\dots,\rho(s=n-2)\}$.
\item Pick $\rho(s)$ in $\{\sigma(2),\dots,\sigma(n-1)\}$. Construct a chain $\mathbb{C}[1]\equiv\{l=1,i_1,\dots,i_j,\rho(s)\}$ of even length started by $\rho(s)$ towards $1$ with internal nodes $i_1, i_2, \dots, i_{j}$ ($j$ is odd) s.t. $\sigma^{-1}(l=1)<\sigma^{-1}({i_1})<\sigma^{-1}({i_2})< \dots < \sigma^{-1}(i_j)<\sigma^{-1}(\rho(s))$. This chain is associated with a factor
    \bea
    s_{\rho(s)i_j}s_{i_{j}i_{j-1}}\dots s_{i_2i_1}s_{i_11}.
    \eea
    Remove this chain from the ordered set $\mathsf{R}\to \mathsf{R}'=\mathsf{R}\setminus \{i_1,i_2, \dots, i_{j},\rho(s)\}\equiv\{\rho'(1),\dots,\rho'(s')\}$.
\item  Repeat the previous step: Pick $\rho'(s')\in \mathsf{R}'$ and construct a chain $\mathbb{C}[2]\equiv\{l',i'_1,\dots,i'_{j'},\rho'({s'})\}$ of even length ($j'$ is odd), which
     starts from $\rho'(s')$ towards a node $l'$ on $\mathbb{C}[1]$ and satisfies
    $\sigma^{-1}(l')<\sigma^{-1}({i'_1})< \dots < \sigma^{-1}(i'_{j'})<\sigma^{-1}(\rho'(s'))$. The new chain $\mathbb{C}[2]$ is associated with a factor
    \bea
    s_{\rho'(s')i'_{j'}}s_{i'_{j'}i'_{j'-1}}\dots s_{i'_2i'_1}s_{i'_1l'}.
    \eea
     Remove this chain from the ordered set $\mathsf{R}\to \mathsf{R}''=\mathsf{R}'\setminus \{i'_1,i'_2, \dots, i'_{j},\rho'(s')\}\equiv\{\rho''(1),\dots,\rho''(s'')\}$.
\item Repeat the above steps until the ordered set $\mathsf{R}$ becomes empty. Each new even-length chain is attached to nodes which have been used and associated with a factor. Collecting the factors corresponding to all chains in a graph and summing over all possible graphs (noting that the total phase factor is $(-1)^{{n\over 2}-1}$), we finally get the BCJ numerator $n^{\text{DT}}_{1|\pmb{\sigma}|n}$.
\end{itemize}
By means of the conventions of notations established for the gauge invariance induced relation \eqref{Eq:NewGaugeIDAmp1}, we can write the numerators of DT type as\footnote{The prefactor  $(-1)^{n-2 \over 2}$ is adjusted by $(-1)$ to agree with that in CMS type. This adjustment does not affect our discussions.}
\bea
n^{\text{DT}}_{1|\pmb{\sigma}|n}=(-1)^{n\over 2}\Sl_{\mathcal{F}\in{\mathcal{G}'^{\,\pmb{\sigma}}_{\{2,\dots,n-1\}}[\emptyset]},
    }\mathcal{D}^{[\mathcal{F}]}(1,\pmb{\sigma},n)\Label{Eq:DTform}
\eea
where the $\mathsf{H}$ set, whose elements serve as starting points or internal nodes of trees, is chosen as $\{2,\dots,n-1\}$. The empty set $\emptyset$ in $\mathcal{G}'^{\,\pmb{\sigma}}_{\{2,\dots,n-1\}}[\emptyset]$ means that all chains are of even length. The explicit expressions for four-point numerators $n^{\text{DT}}_{1|23|4}$ and $n^{\text{DT}}_{1|32|4}$ are given by
\bea
n^{\text{DT}}_{1|23|4}=-s_{32}s_{21},~~~~~~n^{\text{DT}}_{1|32|4}=0,
\eea
where the reference order is chosen as $\mathsf{R}=\{2,3\}$.

\section{The equivalence between DT and CMS constructions of NLSM amplitudes}\label{Sec:DTCMS}
The DT and the CMS  types of numerators produce the same amplitude if and only if the second equality in \eqref{Eq:Equivalence} holds. Substituting  \eqref{Eq:DTform} and \eqref{Eq:CMSform} into  \eqref{Eq:Equivalence},  we arrive the following relation for bi-scalar amplitudes $A(1,\pmb{\sigma},n)$
\bea
\boxed{\Sl_{\pmb{\sigma}\in S_{n-2}}S[\pmb{\sigma}|\pmb{\sigma}]A(1,\pmb{\sigma},n)=\Sl_{\pmb{\sigma}\in S_{n-2}}\Sl_{\mathcal{F}\in{\mathcal{G}'^{\pmb{\sigma}}_{\{2,\dots,n-1\}}[\emptyset]},
    }\mathcal{D}^{[\mathcal{F}]}(1,\pmb{\sigma},n) A(1,\pmb{\sigma},n).~~(\text{for even $n$}) \Label{Eq:EquivDTCMS}}
\eea
To prove the equivalence condition \eqref{Eq:EquivDTCMS}, we carry on our discussions in a more generic framework:
\begin{itemize}
\item [(i)]The momentum kernel $S[\pmb{\sigma}|\pmb{\sigma}]$ is generalized to the partial momentum kernel \bea\W S_{\mathsf{H}}\left[\{2,\dots,r-1\}\shuffle\pmb{\sigma}_{\mathsf{H}}|2,\dots,r-1,\pmb{\sigma}_{\mathsf{H}}\right]\eea where $\mathsf{H}$ is an arbitrary nonempty set with $s$ elements. When setting
    $\{2,\dots,r-1\}=\emptyset$  and $\mathsf{H}=\{2,\dots,n-1\}$, we return to the original momentum kernel $S[\pmb{\sigma}|\pmb{\sigma}]$ ($\pmb{\sigma}\in S_{n-2}$).
\item [(ii)] The number of external particles is not limited to be even. Amplitudes with odd number external particles are also under consideration.
\item [(iii)] The amplitude $A(1,\pmb{\sigma},n)$  can be  color-ordered Yang-Mills, bi-scalar or color-ordered NLSM amplitudes.
\end{itemize}
Having the above generalizations, we will prove the following two relations
\bea
\boxed{\Sl_{\pmb{\sigma}_{\mathsf{H}}}\Sl_{\substack{\pmb{\alpha}\,\in\,\{2,\dots,r-1\}\\\,\,\shuffle\,\pmb{\sigma}_{\mathsf{H}}}}\W S_{\mathsf{H}}\left[\pmb{\alpha}|2,\dots,r-1,\pmb{\sigma}_{\mathsf{H}}\right]A(1,\pmb{\alpha},r)=\Sl_{\substack{\pmb{\alpha}\in\{2,\dots,r-1\}\\\,\shuffle\text{perms}\,{\mathsf{H}}}}\Sl_{\mathcal{F}\in{\mathcal{G}'^{\,\pmb{\alpha}}_{\mathsf{H}}[\emptyset]}
    }\mathcal{D}^{[\mathcal{F}]}(1,\pmb{\alpha},r)A(1,\pmb{\alpha},r)\,(\text{for even $s$})}\Label{Eq:GenEquiv1}\nn
\eea
and
\bea
\boxed{\Sl_{\pmb{\sigma}_{\mathsf{H}}}\Sl_{\substack{\pmb{\alpha}\,\in\,\{2,\dots,r-1\}\\\,\,\shuffle\,\pmb{\sigma}_{\mathsf{H}}}}\W S_{\mathsf{H}}\left[\pmb{\alpha}|2,\dots,r-1,\pmb{\sigma}_{\mathsf{H}}\right]A(1,\pmb{\alpha},r)=0\,(\text{for odd $s$})}\Label{Eq:GenEquiv2}
\eea
corresponding to whether the number of elements in the set $\mathsf{H}$ is even or odd. Coefficients of amplitudes therein are expressed by partial momentum kernels, while the summation $\Sl_{\pmb{\sigma}_{\mathsf{H}}}$ means that we sum over all possible permutations of elements in $\mathsf{H}$. Consequences of the  relations \eqref{Eq:GenEquiv1} and \eqref{Eq:GenEquiv2} are deduced:
\begin{itemize}
 \item When we we set $r=n$ (for even $n$), $\mathsf{H}=\{2,\dots,n-1\}$ and $\{2,\dots,r-1\}\to\emptyset$, the relation \eqref{Eq:GenEquiv1} naturally returns to the equivalence condition \eqref{Eq:EquivDTCMS} for even $n$. Thus the equivalence condition \eqref{Eq:EquivDTCMS} between DT and CMS constructions  is proven.

\item When we set $\{2,\dots,r-1\}\to\{\rho(2),\dots,\rho(r-1)\}$ in the partial momentum kernel  $\W S_{\mathsf{H}}$ in \eqref{Eq:GenEquiv2} and apply the property \eqref{Eq:prop3} and \eqref{Eq:prop1}, the relation \eqref{Eq:GenEquiv2} then becomes
\bea
\Sl_{\pmb{\sigma}_{\mathsf{H}}}\Sl_{\pmb{\alpha}\,\in\,\pmb{\rho}\,\shuffle\,\pmb{\sigma}_{\mathsf{H}}}\W S_{\mathsf{H}}\left[\pmb{\alpha}|2,\dots,r-1,\pmb{\sigma}_{\mathsf{H}}\right]A(1,\pmb{\alpha},r)=0\,(\text{for odd $s$}).
\eea
Multiplying a momentum kernel $S[\rho(2),\rho(3),\dots,\rho(r-1)|2,3,\dots,r-1]$ to both sides of the above relation  and {applying} the relation \eqref{Eq:prop4} between usual momentum kernel and partial momentum kernel, we arrive an amplitude relation expressed by usual momentum kernels
\bea
\boxed{\Sl_{\pmb{\sigma}_{\mathsf{H}}}\Sl_{\substack{\pmb{\alpha}\,\in\,\pmb{\rho}\,\shuffle\,\pmb{\sigma}_{\mathsf{H}}}} S\left[\pmb{\alpha}|2,\dots,r-1,\pmb{\sigma}_{\mathsf{H}}\right]A(1,\pmb{\alpha},r)=0\,(\text{for odd $s$})},\Label{Eq:GenEquiv2-1}
\eea
where $\pmb{\rho}$ is an arbitrary permutation of elements in $\{2,\dots,r-1\}$.
The boundary case with $\mathsf{H}=\{2,\dots,n-1\}$, $\{1,\dots,r\}\to \{1,n\}$ shows very interesting relation for amplitudes with odd number of external particles
\bea
\boxed{\Sl_{\pmb{\sigma}\in S_{n-2}} S\left[\pmb{\sigma}|\pmb{\sigma}\right]A(1,\pmb{\sigma},n)=0\,(\text{for odd $n$})}.\Label{Eq:EquivDTCMS2}
\eea
\end{itemize}
Although the relation \eqref{Eq:GenEquiv2} for odd $s$ is not used in the proof of the equivalence condition \eqref{Eq:EquivDTCMS} between the DT  and the CMS constructions of NLSM amplitudes, the relation \eqref{Eq:GenEquiv2-1} as a result of \eqref{Eq:GenEquiv2}, plays a crucial role in the proof of the equivalence between the DF  and CMS constructions in the next section. In the remaining discussions of this section, we establish the graphic expansion of the partial momentum kernel $\W S_{\mathsf{H}}[\{2,\dots,r-1\}\shuffle\pmb{\sigma}_{\mathsf{H}}|2,\dots,r-1,\pmb{\sigma}_{\mathsf{H}}]$ and prove the relations \eqref{Eq:GenEquiv1} and \eqref{Eq:GenEquiv2}.

\subsection{Expressing partial momentum kernel by graphs}
The partial momentum kernel $\W S_{\mathsf{H}}[\pmb{\alpha}\in\{2,\dots,r-1\}\shuffle\pmb{\sigma}_{\mathsf{H}}|2,\dots,r-1,\pmb{\sigma}_{\mathsf{H}}]$ can be conveniently expanded by the graphic rule in section \eqref{Sec:Expansion}, when replacing the factors $\epsilon_{h_a}\cdot F_{h_{i_1}}\cdot\dots\cdot F_{h_{i_j}}\cdot k_b$ for each chain by $s_{h_ah_{i_1}}s_{h_{i_1}h_{i_2}}\dots s_{h_{i_j}b}$. The reference order $\mathsf{R}=\{h_{\rho(1)},h_{\rho(2)},\dots,h_{\rho(s)}\}$ is chosen arbitrarily. We demonstrate this expansion by examples first.

\subsubsection*{Example-1: $\mathsf{H}=\{h_1,h_2\}$}

The $\sigma_{\mathsf{H}}$ in the partial momentum kernel $\W S_{\{h_1,h_2\}}[\pmb{\alpha}\in\{2,\dots,r-1\}\shuffle \sigma_{\mathsf{H}}|2,\dots,r-1,\sigma_{\mathsf{H}}]$ can be either $\{h_1,h_2\}$ or $\{h_2,h_1\}$. If we define reference order $\mathsf{R}=\{h_1,h_2\}$, the partial momentum kernel with $\sigma_{\mathsf{H}}=\{h_1,h_2\}$ is expressed by the sum of $(a)$ and $(b)$ in figure \ref{Fig:Figure1}, while the partial momentum kernel with $\sigma_{\mathsf{H}}=\{h_2,h_1\}$ is expressed by the sum of $(c)$ and $(d)$ in figure \ref{Fig:Figure1}. If we change the reference order to $\mathsf{R}=\{h_2,h_1\}$, graphs contributing to $\pmb{\sigma}_{\mathsf{H}}=\{h_1,h_2\}$ ($\pmb{\sigma}_{\mathsf{H}}=\{h_2,h_1\}$) become the graphs $(c)$ and $(d)$ ($(a)$ and $(b)$)  in figure \ref{Fig:Figure1} with exchanging $h_1$ and $h_2$. Though the chain structures are different for different choices of reference order, the expression of each partial momentum kernel $\W S_{\{h_1,h_2\}}[\pmb{\alpha}\in\{2,\dots,r-1\}\shuffle \sigma_{\mathsf{H}}|2,\dots,r-1,\sigma_{\mathsf{H}}]$ is not changed.

\subsubsection*{Example-2: $\mathsf{H}=\{h_1,h_2,h_3\}$}

 We now consider the partial momentum kernel
\bea
\W S_{\{h_1,h_2,h_3\}}[\pmb{\alpha}\in\{2,\dots,r-1\}\shuffle\{h_1,h_3,h_2\}|2,\dots,r-1,\{h_1,h_3,h_2\}]\Label{Eq:PartialKerGraphExample}
\eea
where $\mathsf{H}$ contains three elements {and $\pmb{\sigma}_{\mathsf{H}}$ in this example is chosen as $\pmb{\sigma}_{\mathsf{H}}=\{h_1,h_3,h_2\}$}. From the definition \eqref{Eq:PartialMomentumKernal}, \eqref{Eq:PartialKerGraphExample} is given by the product of three factors
\bea
\Biggl[s_{h_11}+\Sl_{\small\substack{i\in\{2,\dots,r-1\}\\ \alpha^{-1}(i)<\alpha^{-1}(h_1) }}s_{h_1i}\Biggr]\Biggl[s_{h_31}+s_{h_3h_1}+\Sl_{\small\substack{i\in\{2,\dots,r-1\}\\ \alpha^{-1}(i)<\alpha^{-1}(h_3) }}s_{h_3i}\Biggr]\Biggl[s_{h_21}+s_{h_2h_1}+s_{h_2h_3}+\Sl_{\small\substack{i\in\{2,\dots,r-1\}\\\alpha^{-1}(i)<\alpha^{-1}(h_2) }}s_{h_2i}\Biggr].\Label{Eq:PartialKerGraphExample1}\nn
\eea
This partial momentum kernel can be obtained as follows:
\begin{itemize}
\item Define a reference order of elements in $\mathsf{H}$, e.g., $\mathsf{R}=\{h_1,h_2,h_3\}$.
\item Pick the last element $h_3$ in the ordered set $\mathsf{R}=\{h_1,h_2,h_3\}$ and pick a term from the factor corresponding to $h_3$ in \eqref{Eq:PartialKerGraphExample1}. Such a term has the form $s_{h_3j}$, where $j$ can be any element in $\{h_1\}\cup\{1,2,\dots,r-1\}$ s.t., $\alpha^{-1}(j)<\alpha^{-1}(h_3)$. If $j$ is an element in $\{1,2,\dots,r-1\}$, we get a length-1 chain started from $h_3$ towards $\{1,2,\dots,r-1\}$. Else, if $j=h_1$, we further pick a factor $s_{h_1k}$ for $k\in\{1,2,\dots,r-1\}$ satisfying $\alpha^{-1}(k)<\alpha^{-1}(h_1)$, then a chain $s_{h_3h_1}s_{h_1k}$ started from $h_3$ towards $k$ have been constructed. We take the $j=h_1$ case for instance and continue our discussion.

\item Remove the starting node $h_3$ and the internal node $h_1$ of the chain which have been already constructed, from the ordered set $\mathsf{R}=\{h_1,h_2,h_3\}$ and redefine $\mathsf{R}$ as  $\mathsf{R}\to\mathsf{R}'=\{h_2\}$. Construct a chain started from the element $h_2$ in $\mathsf{R}'$ towards $l\in\{h_1,h_3\}\cup\{1,2,\dots,r-1\}$. Then we have a factor $s_{h_2l}$. For example, we choose $l=h_1$.

\item Remove $h_2$ from $\mathsf{R}'$, then the set $\mathsf{R}'$ becomes empty. Putting the chains obtained together, we arrive a term $s_{h_3h_1}s_{h_1k}s_{h_2h_1}$ corresponding to the graph $(b4)$ of figure \ref{Fig:5ptGraphs}.
\item The full partial momentum kernel in this example is obtained by summing over all possible graphs constructed by the above steps (displayed by the graphs $(b1)\sim(b6)$ in figure \ref{Fig:5ptGraphs}).
\end{itemize}
Again, we emphasize that the reference order $\mathsf{R}$ can be chosen arbitrarily. If we change the reference order, only the chains are changed, the structure of graphs and the final expression of partial momentum kernel are not changed.

Now we extend our discussions to the graphic expansion of any partial momentum kernel with the form:
\bea
&&\W S_{\mathsf{H}}[\{2,\dots,r-1\}\shuffle\pmb{\sigma}_{\mathsf{H}}|2,\dots,r-1,\pmb{\sigma}_{\mathsf{H}}]\nn
&=&\Biggl[s_{{\sigma}_{\mathsf{H}}(1)1}+\Sl_{\small \substack{i\in\{2,\dots,r-1\}\\ \alpha^{-1}(i)<\alpha^{-1}(\sigma_{{\mathsf{H}}(1)})}}s_{{\sigma}_{\mathsf{H}}(1)i}\Biggr]\Biggl[s_{{\sigma}_{\mathsf{H}}(2)1}+s_{{\sigma}_{\mathsf{H}}(2)\sigma_{\mathsf{H}}(1)}+\Sl_{\small\substack{i\in\{2,\dots,r-1\}\\ \alpha^{-1}(i)<\alpha^{-1}(\sigma_{{\mathsf{H}}(2)})}}s_{{\sigma}_{\mathsf{H}}(2)i}\Biggr]\nn
&&\times\dots\times\Biggl[s_{{\sigma}_{\mathsf{H}}(s)1}+s_{{\sigma}_{\mathsf{H}}(s)\sigma_{\mathsf{H}}(1)}+\dots s_{{\sigma}_{\mathsf{H}}(s)\sigma_{\mathsf{H}}(s-1)}+\Sl_{\small\substack{i\in\{2,\dots,r-1\}\\ \alpha^{-1}(i)<\alpha^{-1}(\sigma_{{\mathsf{H}}(s)})}}s_{{\sigma}_{\mathsf{H}}(s)i}\Biggr].\Label{Eq:PartialKerGraph}
\eea
\begin{itemize}
\item Define a reference order  $\mathsf{R}=\{h_{\rho(1)},h_{\rho(2)},\dots,h_{\rho(s)}\}$ for elements in the set $\mathsf{H}$ (assume there are $s$ elements in the set $\mathsf{H}$). Pick $h_{\rho(s)}$ and an arbitrary term  $s_{h_{\rho(s)h_{i_j}}}$ ($\sigma^{-1}(h_{i_j})<\sigma^{-1}(h_{\rho(s)})$) from the factor corresponding to $h_{\rho(s)}$. Then pick an arbitrary term $s_{h_{i_j}h_{i_{j-1}}}$ ($\sigma^{-1}(h_{i_{j-1}})<\sigma^{-1}(h_{i_j})$) from the factor corresponding to $h_{i_j}$. Next, pick a term of the form $s_{h_{i_{j-1}}h_{i_{j-2}}}$ ($\sigma^{-1}(h_{i_{j-2}})<\sigma^{-1}(h_{i_{j-1}})$) from the factor corresponding to $h_{i_{j-1}}$, and so on. This procedure is terminated at a factor $s_{h_{i_1}l}$ where $l$ belongs to the set $\{1,2,\dots,r-1\}$. Putting all factors together, we get a chain $s_{h_{\rho(s)h_{i_j}}}s_{h_{i_j}h_{i_{j-1}}}\dots s_{h_{i_1}l}$. Redefine $\mathsf{R}$ by removing the internal nodes and the starting point of the chain which was already constructed: $\mathsf{R}\to \mathsf{R}'=\mathsf{R}\setminus \{h_{i_1},\dots,h_{i_j},h_{\rho(s)}\}\equiv \{h_{\rho'(1)},h_{\rho'(2)},\dots,h_{\rho'(s')}\}$.

\item We construct a chain from $h_{\rho'(s')}$ towards an element {$l'\in\{1,2,\dots,r\}\cup\{h_{i_1},\dots,h_{i_j},h_{\rho(s)}\}$} by picking $s_{h_{\rho'(s')}h_{i'_{j'}}}$, $s_{h_{i'_{j'}}h_{i'_{j'-1}}}$, ...,  $s_{h_{i'_{1}}l'}$ ($\sigma^{-1}(l')<\sigma^{-1}(h_{i'_{1}})<\dots<\sigma^{-1}(h_{i'_{j'}})<\sigma^{-1}(h_{\rho(s')})$) from the factors corresponding to $h_{\rho(s')}$, $h_{i'_{j'}}$, ..., $h_{i'_{1}}$ in the partial momentum kernel \eqref{Eq:PartialKerGraph}. The we get another chain $s_{h_{\rho'(s')}h_{i'_{j'}}}s_{h_{i'_{j'}}h_{i'_{j'-1}}}\dots s_{h_{i'_{1}}l'}$. Redefine $\mathsf{R}$ by  $\mathsf{R}\to \mathsf{R}''=\mathsf{R}'\setminus \{h_{i'_{1}},\dots,h_{i'_{j'}},h_{\rho'(s')}\}$.

\item  Repeat the above steps until the $\mathsf{R}$ set becomes empty. Then putting all chains together, we get a graph. The sum of all possible graphs gives the partial momentum kernel \eqref{Eq:PartialKerGraph}.
\end{itemize}
Obviously, if we define a unique reference order $\mathsf{R}$ for permutations $\pmb{\alpha}\in\{2,\dots,r-1\}\shuffle\pmb{\sigma}_{\mathsf{H}}$ with all possible $\pmb{\sigma}_{\mathsf{H}}$, the above graphic expansions of partial momentum kernels $\W S_{\mathsf{H}}[\pmb{\alpha}\in\{2,\dots,r-1\}\shuffle\pmb{\sigma}_{\mathsf{H}}|2,\dots,r-1,\pmb{\sigma}_{\mathsf{H}}]$ are related with the graphic expansion of $\mathcal{C}(1,\pmb{\sigma},r)$ (see \eqref{Eq:Coefficients})
in section \eqref{Sec:Expansion} via replacing the factor $\epsilon_{h_a}\cdot F_{h_{i_j}}\cdot\dots\cdot F_{h_{i_1}}\cdot k_b$  for every chain by $s_{h_ah_{i_j}}s_{h_{i_j}h_{i_{j-1}}}\dots s_{h_{i_1}b}$.

\subsection{Proof of the relations \eqref{Eq:GenEquiv1} and \eqref{Eq:GenEquiv2}}
We have already shown that the equivalence condition \eqref{Eq:EquivDTCMS} is a special case of the relation \eqref{Eq:GenEquiv1} with even $s$.
In addition, we also have the relation \eqref{Eq:GenEquiv2} with odd $s$. Now let us prove both relations \eqref{Eq:GenEquiv1} and \eqref{Eq:GenEquiv2} by expanding the partial momentum kernels into graphs.


\subsubsection{The proof of \eqref{Eq:GenEquiv1}}
To prove the relation \eqref{Eq:GenEquiv1} for even $s$, we first investigate two examples.

{\bf Example-1: $\mathsf{H}=\{h_1,h_2\}$}~~The simplest example for even $s$ is the case $\mathsf{H}=\{h_1,h_2\}$ (hence $s=2$). If we choose reference order as $\mathsf{R}=\{h_1,h_2\}$, the graphs corresponding to $\sigma_{\mathsf{H}}=\{h_1,h_2\}$ ($\sigma_{\mathsf{H}}=\{h_2,h_1\}$) are explicitly given by $(a)$ and $(b)$ ($(c)$ and $(d)$)in figure \ref{Fig:Figure1}. The LHS of \eqref{Eq:GenEquiv1} for this case reads
\bea
&&\Sl_{\pmb{\alpha}\in\{2,\dots,r-1\}\shuffle\{h_1,h_2\}}\W S_{\{h_1,h_2\}}\bigl[\pmb{\alpha}\big|2,\dots,r-1,h_1,h_2\bigr]A(1,\pmb{\alpha},r)\nn
&+&\Sl_{\pmb{\alpha}\in\{2,\dots,r-1\}\shuffle\{h_2,h_1\}}\W S_{\{h_1,h_2\}}\bigl[\pmb{\alpha}\big|2,\dots,r-1,h_2,h_1\bigr]A(1,\pmb{\alpha},r).
\eea
Expanding the partial momentum kernels into graphs (see figure \ref{Fig:Figure1}), we rewrite the above expression as
\bea
&&\Sl_{\pmb{\alpha}\in\{2,\dots,r-1\}\shuffle\{h_1,h_2\}}\Bigl[\mathcal{D}^{[(a)]}(1,\pmb{\alpha},r)+\mathcal{D}^{[(b)]}(1,\pmb{\alpha},r)\Bigr]A(1,\pmb{\alpha},r)\nn
&+&\Sl_{\pmb{\alpha}\in\{2,\dots,r-1\}\shuffle\{h_2,h_1\}}\Bigl[\mathcal{D}^{[(c)]}(1,\pmb{\alpha},r)+\mathcal{D}^{[(d)]}(1,\pmb{\alpha},r)\Bigr]A(1,\pmb{\alpha},r),
\eea
where $\mathcal{D}^{[(a)]}(1,\pmb{\alpha},r)$, $\mathcal{D}^{[(b)]}(1,\pmb{\alpha},r)$, $\mathcal{D}^{[(c)]}(1,\pmb{\alpha},r)$ and $\mathcal{D}^{[(d)]}(1,\pmb{\alpha},r)$ are coefficients associating to the graphs $(a)$, $(b)$, $(c)$ and $(d)$ in figure \ref{Fig:Figure1}. The graphs $(a)$, $(c)$ and $(d)$ in the above equation contain two length-1 chains. They together contribute
\bea
&&\Sl_{\pmb{\alpha}\in\{2,\dots,r-1\}\shuffle\{h_1,h_2\}}\mathcal{D}^{[(c)]}(1,\pmb{\alpha},r)A(1,\pmb{\alpha},r)+\Sl_{\pmb{\alpha}\in\{2,\dots,r-1\}\shuffle\{h_2,h_1\}}\Bigl[\mathcal{D}^{[(c)]}(1,\pmb{\alpha},r)+\mathcal{D}^{[(d)]}(1,\pmb{\alpha},r)\Bigr]A(1,\pmb{\alpha},r)\nn
&=&\Sl_{\pmb{\alpha}\in\{2,\dots,r-1\}\shuffle\,\text{perms}\,\{h_1,h_2\}}\Sl_{\mathcal{F}\in{\mathcal{G}'^{\pmb{\alpha}}_{\{h_1,h_2\}}[\{h_1,h_2\}]}
    }\,\,\mathcal{D}^{[\mathcal{F}]}(1,\pmb{\alpha},r)A(1,\pmb{\alpha},r),
\eea
which is nothing but the RHS of the example \eqref{Eq:GaugeInducedExample1}, thus have to vanish. The only term that survives is the graph $(b)$ which contains no odd length chain
\bea
\Sl_{\pmb{\alpha}\in\{2,\dots,r-1\}\shuffle\{h_1,h_2\}}\mathcal{D}^{[(b)]}(1,\pmb{\alpha},r)A(1,\pmb{\alpha},r)=\Sl_{\pmb{\alpha}\in\{2,\dots,r-1\}\shuffle\,\text{perms}\,\{h_1,h_2\}}\Sl_{\mathcal{F}\in{\mathcal{G}'^{\pmb{\alpha}}_{\{h_1,h_2\}}[\emptyset]}
    }\,\,\mathcal{D}^{[\mathcal{F}]}(1,\pmb{\alpha},r)A(1,\pmb{\alpha},r),\nn
\eea
agrees with the RHS of \eqref{Eq:GenEquiv1} with $\mathsf{H}=\{h_1,h_2\}$.

{\bf Example-2: $\mathsf{H}=\{h_1,h_2,h_3,h_4\}$}~~Inspired by the previous example with $s=2$, one can expand all partial momentum kernels on the LHS of \eqref{Eq:GenEquiv1} in terms of graphs for a given reference order $\mathsf{R}$. For the case $\mathsf{H}=\{h_1,h_2,h_3,h_4\}$, the total length of all chains of each expansion graph should equal to $4$. On the other hand, the total length $L^{\text{total}}$ of all chains is given by
\bea
L^{\text{total}}=L^{\text{odd}}+L^{\text{even}},
\eea
where $L^{\text{odd}}$ and $L^{\text{even}}$ denote the total lengths of all odd- and even-length chains, respectively. If a graph contains odd number of odd-length chains, the total length must be odd according to the above equation. This conflicts with the fact $L^{\text{total}}=4$. Therefore, the number of odd-length chains must be even. In this example, each graph can contain 0, 2 or 4 odd-length chains. Thus for $\mathsf{H}=\{h_1,h_2,h_3,h_4\}$, the LHS of  \eqref{Eq:GenEquiv1} is expanded as
\bea
&&\Sl_{\pmb{\alpha}\in\{2,\dots,r-1\}\shuffle\,\text{perms}\,\mathsf{H}}\,\,\,\,\,\,\,\,\,\Sl_{\mathcal{F}\in{\mathcal{G}'^{\pmb{\alpha}}_{\mathsf{H}}[\emptyset]}
    }\mathcal{D}^{[\mathcal{F}]}(1,\pmb{\alpha},r)A(1,\pmb{\alpha},r)\nn
    &+&\Sl_{\pmb{\alpha}\in\{2,\dots,r-1\}\shuffle\,\text{perms}\,\mathsf{H}}\biggl[\Sl_{\mathcal{F}\in{\mathcal{G}'^{\pmb{\alpha}}_{\mathsf{H}}[\{h_1,h_2\}]}}
    +\Sl_{\mathcal{F}\in{\mathcal{G}'^{\pmb{\alpha}}_{\mathsf{H}}[\{h_1,h_3\}]}}+\Sl_{\mathcal{F}\in{\mathcal{G}'^{\pmb{\alpha}}_{\mathsf{H}}[\{h_1,h_4\}]}}\nn
    &&~~~~~~~~~~~~~~~~~~~~~~+\Sl_{\mathcal{F}\in{\mathcal{G}'^{\pmb{\alpha}}_{\mathsf{H}}[\{h_2,h_3\}]}}
    +\Sl_{\mathcal{F}\in{\mathcal{G}'^{\pmb{\alpha}}_{\mathsf{H}}[\{h_2,h_4\}]}}+\Sl_{\mathcal{F}\in{\mathcal{G}'^{\pmb{\alpha}}_{\mathsf{H}}[\{h_3,h_4\}]}}\biggr]\mathcal{D}^{[\mathcal{F}]}(1,\pmb{\alpha},r)A(1,\pmb{\alpha},r)\nn
    &+&\Sl_{\pmb{\alpha}\in\{2,\dots,r-1\}\shuffle\,\text{perms}\,\mathsf{H}}\Sl_{\mathcal{F}\in{\mathcal{G}'^{\pmb{\alpha}}_{\mathsf{H}}[\{h_1,h_2,h_3,h_4\}]}
    }\mathcal{D}^{[\mathcal{F}]}(1,\pmb{\alpha},r)A(1,\pmb{\alpha},r).
\eea
The last three lines vanishes due to the gauge invariance induced relation \eqref{Eq:NewGaugeIDAmp1} with $\mathsf{A}=\{h_i,h_j\}$ ($h_i,h_j\in \mathsf{H}$) and $\mathsf{A}=\{h_1, h_2, h_3, h_4\}$ (the case with $\mathsf{A}=\{h_3,h_4\}$ and $\mathsf{R}=\{h_1,h_2,h_3,h_4\}$ is explicitly given by the example \eqref{Eq:GaugeInducedExample6}), while the first line is the RHS of \eqref{Eq:GenEquiv1} for $s=4$.

{\bf General proof of \eqref{Eq:GenEquiv1}}~~If $\mathsf{H}$ contains an arbitrary even number of elements ({\it i.e.}, $s$ is even), the number of odd-length chains in any graph has to be even, as analyzed in the $s=4$ example. Thus the partial momentum kernel can be written as
\bea
&&S_{\mathsf{H}}\Bigl[\pmb{\alpha}\in\{2,\dots,r-1\}\shuffle\pmb{\sigma}_{\mathsf{H}}\Big|2,\dots,r-1,\pmb{\sigma}_{\mathsf{H}}\Bigr]\\
&=&\Sl_{\mathcal{F}\in{\mathcal{G}'^{\pmb{\alpha}}_{\mathsf{H}}[\emptyset]}
    }\mathcal{D}^{[\mathcal{F}]}(1,\pmb{\alpha},r)+\Sl_{\{h_{i_1},h_{i_2}\}\subset\mathsf{H}}\Sl_{\mathcal{F}\in{\mathcal{G}'^{\pmb{\alpha}}_{\mathsf{H}}[\{h_{i_1},h_{i_2}\}]}
    }\mathcal{D}^{[\mathcal{F}]}(1,\pmb{\alpha},r)+\dots+\Sl_{\mathcal{F}\in{\mathcal{G}'^{\pmb{\alpha}}_{\mathsf{H}}[\mathsf{H}]}
    }\mathcal{D}^{[\mathcal{F}]}(1,\pmb{\alpha},r).~~\text{(for even $s$)}\nonumber
\eea
Then the combination of amplitudes on the LHS of \eqref{Eq:GenEquiv1}  turns to
\bea
&&\Sl_{\substack{\pmb{\alpha}\,\in\,\{2,\dots,r-1\}\\\,\,\shuffle\,\text{perms\,}{\mathsf{H}}}} \Bigl[\Sl_{\mathcal{F}\in{\mathcal{G}'^{\pmb{\alpha}}_{\mathsf{H}}[\emptyset]}
    }\mathcal{D}^{[\mathcal{F}]}(1,\pmb{\alpha},r)+\Sl_{\{h_{i_1},h_{i_2}\}\subset\mathsf{H}}\Sl_{\mathcal{F}\in{\mathcal{G}'^{\pmb{\alpha}}_{\mathsf{H}}[\{h_{i_1},h_{i_2}\}]}
    }\mathcal{D}^{[\mathcal{F}]}(1,\pmb{\alpha},r)\nn
&&~~~~~~~~~~~~~~+\dots+\Sl_{\mathcal{F}\in{\mathcal{G}'^{\pmb{\alpha}}_{\mathsf{H}}[\mathsf{H}]}
    }\mathcal{D}^{[\mathcal{F}]}(1,\pmb{\alpha},r)\Bigr]A(1,\pmb{\alpha},r)\Label{Eq:PartialKerGraph1}
\eea
Every term in the above expression  have the general form
\bea
\Sl_{\substack{\pmb{\alpha}\,\in\,\{2,\dots,r-1\}\\\,\,\shuffle\,\text{perms}\, \mathsf{H}}} \Sl_{\mathcal{F}\in{\mathcal{G}'^{\pmb{\alpha}}_{\mathsf{H}}[\{h_{i_1}h_{i_2},\dots,h_{i_j}\}]}
    }\mathcal{D}^{[\mathcal{F}]}(1,\pmb{\alpha},r)A(1,\pmb{\alpha},r).~~\text{(for even $s$ and $j$)}
\eea
If $j\neq 0$, the set  $\{h_{i_1}h_{i_2},\dots,h_{i_j}\}\subseteq \mathsf{H}$ is nonempty. Such a term has to vanish due to the gauge invariance induced relation \eqref{Eq:NewGaugeIDAmp1} for the nonempty subset $\mathsf{A}$ with even number of elements.
The first term in \eqref{Eq:PartialKerGraph1} (the case $j=0$) is given by summing over all graphs consisting of only even length chains, which is the RHS of \eqref{Eq:GenEquiv1}.

\subsubsection{The proof of \eqref{Eq:GenEquiv2}}
The first nontrivial example of \eqref{Eq:GenEquiv2} for odd $s$ is given by $\mathsf{H}=\{h_1,h_2,h_3\}$. Let us study this case before the general proof of \eqref{Eq:GenEquiv2}.

{\bf Example: $\mathsf{H}=\{h_1,h_2,h_3\}$}~~We expand the partial momentum kernels on the LHS of \eqref{Eq:GenEquiv2} in terms of graphs for a fixed reference order $\mathsf{R}=\{h_{\rho(1)},h_{\rho(2)},h_{\rho(3)}\}$. For a given graph, the total length of all chains must be $3$. As a consequence, the number of odd length chains in each graph must be odd (in this example it can be $1$ or $3$). Thus the LHS of  \eqref{Eq:GenEquiv2} for $s=3$ is decomposed into
\bea
    &&\Sl_{\pmb{\alpha}\in\{2,\dots,r-1\}\shuffle\,\text{perms}\,\mathsf{H}}\biggl[\Sl_{\mathcal{F}\in{\mathcal{G}'^{\pmb{\alpha}}_{\mathsf{H}}[\{h_1\}]}}
    +\Sl_{\mathcal{F}\in{\mathcal{G}'^{\pmb{\alpha}}_{\mathsf{H}}[\{h_2\}]}}+\Sl_{\mathcal{F}\in{\mathcal{G}'^{\pmb{\alpha}}_{\mathsf{H}}[\{h_3\}]}}\biggr]\mathcal{D}^{[\mathcal{F}]}(1,\pmb{\alpha},r)A(1,\pmb{\alpha},r)\nn
    &+&\Sl_{\pmb{\alpha}\in\{2,\dots,r-1\}\shuffle\,\text{perms}\,\mathsf{H}}\Sl_{\mathcal{F}\in{\mathcal{G}'^{\pmb{\alpha}}_{\mathsf{H}}[\{h_1,h_2,h_3\}]}
    }\mathcal{D}^{[\mathcal{F}]}(1,\pmb{\alpha},r)A(1,\pmb{\alpha},r),
\eea
where each term on the first line vanishes due to the gauge invariance  induced relation \eqref{Eq:NewGaugeIDAmp1} with $\mathsf{A}=\{h_i\}$ ($i=1,2,3$) (see the examples \eqref{Eq:GaugeInducedExample2}, \eqref{Eq:GaugeInducedExample3} and \eqref{Eq:GaugeInducedExample4} for $\mathsf{R}=\{h_{1},h_{2},h_{3}\}$), while the last line vanishes because of the relation \eqref{Eq:NewGaugeIDAmp1} with $\mathsf{A}=\{h_1,h_2,h_3\}$ (see the example \eqref{Eq:GaugeInducedExample5} for $\mathsf{R}=\{h_{1},h_{2},h_{3}\}$).
Hence all terms of the LHS of \eqref{Eq:GenEquiv2} for $s=3$ vanish and the equation \eqref{Eq:GenEquiv2} for $s=3$ is proven.

{\bf General proof of \eqref{Eq:GenEquiv2}}~~
If $\mathsf{H}$ contains an arbitrary odd number of elements ({\it i.e.}, $s$ is odd), the number of odd length chains in any graph must be odd as shown in the $s=3$ example.
The graphic expansions of partial momentum kernels then read
\bea
&&S_{\mathsf{H}}\Bigl[\pmb{\alpha}\in\{2,\dots,r-1\}\shuffle\pmb{\sigma}_{\mathsf{H}}\Big|2,\dots,r-1,\pmb{\sigma}_{\mathsf{H}}\Bigr]\nn
&=&\Sl_{\{h_{i_1}\}\subset\mathsf{H}}\Sl_{\mathcal{F}\in{\mathcal{G}'^{\pmb{\alpha}}_{\mathsf{H}}[\{h_{i_1}\}]}
    }\mathcal{D}^{[\mathcal{F}]}(1,\pmb{\alpha},r)+\Sl_{\{h_{i_1},h_{i_2},h_{i_3}\}\subset \mathsf{H}}\Sl_{\mathcal{F}\in{\mathcal{G}'^{\pmb{\alpha}}_{\mathsf{H}}[\{h_{i_1},h_{i_2},h_{i_3}\}]}
    }\mathcal{D}^{[\mathcal{F}]}(1,\pmb{\alpha},r)+\dots+\Sl_{\mathcal{F}\in{\mathcal{G}'^{\pmb{\alpha}}_{\mathsf{H}}[\mathsf{H}]}
    }\mathcal{D}^{[\mathcal{F}]}(1,\pmb{\alpha},r)\nn
&&~~~~~~~~~~~~~~~~~~~~~~~~~~~~~~~~~~~~~~~~~~~~~~~~~~~~~~~~~~~~~~~~~~~~~~~~~~~~~~~~~~~~~~~~~~~~~~~~~~~~~~~~~~~~~\,\text{(for odd $s$)}.
\eea
The combination of amplitudes in the LHS of \eqref{Eq:GenEquiv2} leads to
\bea
&&\Sl_{\substack{\pmb{\alpha}\,\in\,\{2,\dots,r-1\}\\\,\,\shuffle\,\text{perms\,}\mathsf{H}}} \Bigl[\Sl_{\{h_{i_1}\}\subset \mathsf{H}}\Sl_{\mathcal{F}\in{\mathcal{G}'^{\pmb{\alpha}}_{\mathsf{H}}[\{h_{i_1}\}]}
    }\mathcal{D}^{[\mathcal{F}]}(1,\pmb{\alpha},r)+\Sl_{\{h_{i_1},h_{i_2},h_{i_3}\}\subset \mathsf{H}}\Sl_{\mathcal{F}\in{\mathcal{G}'^{\pmb{\alpha}}_{\mathsf{H}}[\{h_{i_1},h_{i_2},h_{i_3}\}]}
    }\mathcal{D}^{[\mathcal{F}]}(1,\pmb{\alpha},r)\nn
&&~~~~~~~~~~~~~~+\dots+\Sl_{\mathcal{F}\in{\mathcal{G}'^{\pmb{\alpha}}_{\mathsf{H}}[\mathsf{H}]}
    }\mathcal{D}^{[\mathcal{F}]}(1,\pmb{\alpha},r)\Bigr]A(1,\pmb{\alpha},n),
\eea
in which, all terms must vanish due to the gauge invariance induced relation \eqref{Eq:NewGaugeIDAmp1} for $\mathsf{A}$ with odd number of elements. Thus the relation \eqref{Eq:GenEquiv2} is proven.

\section{The equivalence between DF and CMS constructions of NLSM amplitudes}\label{Sec:DFCMS}
The equivalence between DF  and CMS constructions of NLSM amplitudes, {\it i.e.}, the first equality of \eqref{Eq:Equivalence} can be explicitly expressed by the following amplitude relation
\bea
\boxed{\Sl_{\pmb{\sigma}\in S_{n-2}}\Sl_{\pmb{\rho}\in \mathsf{\Gamma}}S[\pmb{\sigma}|\pmb{\rho}]A(1,\pmb{\sigma},n)=(-1)^{n-2\over 2}\Sl_{\pmb{\sigma}\in S_{n-2}}S[\pmb{\sigma}|\pmb{\sigma}]A(1,\pmb{\sigma},n),~~~(\text{for even $n$}) }\Label{Eq:EquivDFCMS}
\eea
where $\mathsf{\Gamma}$ is defined in section 4. In this section, we will prove the relation \eqref{Eq:EquivDFCMS}. {The identity \eqref{Eq:GenEquiv2-1} (as a consequence \eqref{Eq:GenEquiv2})} with odd $s$ is crucial for the proof. To show the pattern, let us first discuss the four- and six-point examples as a warmup.
\subsection{Warm-up examples}
Now we take the cases with $n=4$ and $n=6$ as examples.
\subsubsection*{Four-point example}The simplest example is the four-point case, which have already been discussed in \cite{Carrasco:2016ldy} and \cite{Du:2017kpo}. The LHS of the relation \eqref{Eq:EquivDFCMS} for $n=4$ is explicitly written as
\bea
S[23|32]A(1,2,3,4)+S[32|23]A(1,3,2,4).
\eea
Applying the relation \eqref{Eq:GenEquiv2-1} with $\mathsf{H}=\{2\}$ and $\mathsf{H}=\{3\}$ on the first and the second terms respectively, we immediately get
\bea
-S[32|32]A(1,3,2,4)-S[23|23]A(1,2,3,4),
\eea
which is the RHS of \eqref{Eq:EquivDFCMS} for four-point case.

\subsubsection*{Six-point example}
The relation \eqref{Eq:EquivDFCMS} for six-point amplitudes is much more nontrivial. By substituting the six-point numerators of DF type \eqref{Eq:DF6Pt} into the LHS of  \eqref{Eq:EquivDFCMS}, we get
\bea
&&\Sl_{\pmb{\sigma}\in S_{4}}\Bigl(S\left[\pmb{\sigma}|\pmb{\rho}=\{\sigma(5),\sigma(2),\sigma(4),\sigma(3)\}\right]+S\left[\pmb{\sigma}|\pmb{\rho}=\{\sigma(5),\sigma(3),\sigma(4),\sigma(2)\}\right]\nn
&&~~~\,\,+S\left[\pmb{\sigma}|\pmb{\rho}=\{\sigma(4),\sigma(3),\sigma(5),\sigma(2)\}\right]+S\left[\pmb{\sigma}|\pmb{\rho}=\{\sigma(4),\sigma(2),\sigma(5),\sigma(3)\}\right] \nn
&&~~~\,\,+S\left[\pmb{\sigma}|\pmb{\rho}=\{\sigma(3),\sigma(2),\sigma(5),\sigma(4)\}\right]\Bigr)A(1,\pmb{\sigma},n).\Label{Eq:EquivDFCMS6Pt}
\eea
To prove this expression equals to the RHS of   \eqref{Eq:EquivDFCMS} with $n=6$, we perform our discussions by the following steps.
\newline
{\bf\emph{Step-1}} Collect those terms with a same $\pmb{\rho}$. For example, if $\pmb{\rho}=\{2,3,4,5\}$, one finds that the corresponding $\pmb{\sigma}$ in \eqref{Eq:EquivDFCMS6Pt} can be
\bea
 \{5,3,4,2\},~~~~\{5,3,2,4\},~~~~\{3,5,4,2\},~~~~\{3,5,2,4\},~~~~\{3,2,5,4\}. \Label{Eq:DFCMSPerms6Pt}
\eea
An interesting observation is that above permutations are those satisfying the `zigzag pattern': $\sigma^{-1}(5)<\sigma^{-1}(4)$, $\sigma^{-1}(4)>\sigma^{-1}(3)$ and $\sigma^{-1}(3)<\sigma^{-1}(2)$. For convenience, we define the collection  of such permutations by $\mathsf{Z}\{2|3|4|5\}$:
\bea
\mathsf{Z}\{2|3|4|5\}\equiv\left\{\pmb{\sigma}\,|\,\sigma\in S_4\text{~s.t~} \sigma^{-1}(5)<\sigma^{-1}(4),\,\sigma^{-1}(4)>\sigma^{-1}(3),\,\sigma^{-1}(3)<\sigma^{-1}(2)\right\}.
\eea
Under this definition, terms with $\pmb{\rho}=\{2,3,4,5\}$ in \eqref{Eq:EquivDFCMS6Pt} then give rise
\bea
T(2|3|4|5)\equiv \Sl_{\pmb{\sigma}\in\mathsf{Z}\{2|3|4|5\}}S[\pmb{\sigma}|2,3,4,5]A(1,\pmb{\sigma},6). \Label{Eq:DFCMSExample1}
\eea
Terms corresponding to arbitrary $\pmb{\rho}$ can be obtained by relabeling the above expression
\bea
T(\rho(2)|\rho(3)|\rho(4)|\rho(5))\equiv \Sl_{\pmb{\sigma}\in\mathsf{Z}\{\rho(2)|\rho(3)|\rho(4)|\rho(5)\}}S[\pmb{\sigma}|\pmb{\rho}]A(1,\pmb{\sigma},6).\Label{Eq:DFCMSExampleT0}
\eea
All together, \eqref{Eq:EquivDFCMS6Pt} becomes
\bea
\Sl_{\pmb{\rho}\in S_4}T(\rho(2)|\rho(3)|\rho(4)|\rho(5)).\Label{Eq:DFCMSExampleT1}
\eea
\newline
{\bf\emph{ Step-2}} For a given $\pmb{\rho}$, we collect terms corresponding to those permutations $\pmb{\sigma}$ ($\pmb{\sigma}\in\mathsf{Z}\{\rho(2)|\rho(3)|\rho(4)|\rho(5)\}$)  in which  $\rho(2)$, $\rho(3)$ and $\rho(4)$ have a same relative order. For instance, in the case $\pmb{\rho}=\{2,3,4,5\}$, $T(2|3|4|5)$ then becomes
\bea
T(2|3|4|5)&=&\Bigl[S[5,3,4,2|2,3,4,5]A(1,5,3,4,2,6)+S[3,5,4,2|2,3,4,5]A(1,3,5,4,2,6)\Bigr]\nn
&&+\Bigl[S[5,3,2,4|2,3,4,5]A(1,5,3,2,4,6)+S[3,5,2,4|2,3,4,5]A(1,3,5,2,4,6)\nn
&&~~+S[3,2,5,4|2,3,4,5]A(1,3,2,5,4,6)\Bigr],
\eea
where the first line gets contribution from permutations $\pmb{\sigma}\in\mathsf{Z}\{2|3|4|5\}$ with the relative order $\{3,4,2\}$; the second and the third lines get contributions from $\pmb{\sigma}\in\mathsf{Z}\{2|3|4|5\}$ with the relative order $\{3,2,4\}$. By means of the property \eqref{Eq:GenEquiv2-1} with ($\mathsf{H}=\{5\}$), we write the first line in the above expression as
\bea
-S[3,4,5,2|2,3,4,5]A(1,3,4,5,2,6)-S[3,4,2,5|2,3,4,5]A(1,3,4,2,5,6). \Label{Eq:EquivDFCMS6Pt1}
\eea
Similarly, the second and the third lines sum to
\bea
-S[3,2,4,5|2,3,4,5]A(1,3,2,4,5,6).\Label{Eq:EquivDFCMS6Pt2}
\eea
If we define
\bea
\mathsf{Z}\{2|3,4,5\}\equiv\left\{\pmb{\sigma}\,|\,\sigma\in S_4\text{~s.t~} \sigma^{-1}(3)<\sigma^{-1}(4)<\sigma^{-1}(5),\,\sigma^{-1}(3)<\sigma^{-1}(2)\right\},
\eea
the sum of \eqref{Eq:EquivDFCMS6Pt1} and \eqref{Eq:EquivDFCMS6Pt2} are further expressed by
\bea
T(2|3|4|5)=(-1)T(2|3,4,5)\equiv (-1)\Sl_{\pmb{\sigma}\in\mathsf{Z}\{2|3,4,5\}}S[\pmb{\sigma}|2,3,4,5]A(1,\pmb{\sigma},6).
\eea
For the same reason, $T(\rho(2)|\rho(3)|\rho(4)|\rho(5))$ for arbitrary $\pmb{\rho}$ is written as
\bea
T(\rho(2)|\rho(3)|\rho(4)|\rho(5))&=&(-1)T(\rho(2)|\rho(3),\rho(4),\rho(5))\nn
&\equiv& (-1)\Sl_{\pmb{\sigma}\in\mathsf{Z}\{\rho(2)|\rho(3),\rho(4),\rho(5)\}}S[\pmb{\sigma}|\rho(2),\rho(3),\rho(4),\rho(5)]A(1,\pmb{\sigma},6).
\eea
Therefore \eqref{Eq:DFCMSExampleT1} turns to
\bea
(-1)\Sl_{\pmb{\rho}\in S_4}T(\rho(2)|\rho(3),\rho(4),\rho(5)).\Label{Eq:DFCMSExampleT2}
\eea
\newline
{\bf \emph{Step}-3} Now we collect terms in the combination of amplitudes \eqref{Eq:DFCMSExampleT2} for a given element $\rho(2)\in\{2,3,4,5\}$. In the case of $\rho(2)=2$, we have
\bea
(-1)\Sl_{\pmb{\sigma}\in\text{perms~}\{3,4,5\}}T(2|\pmb{\sigma})=(-1)\Sl_{\pmb{\sigma}\in\text{perms~}\{3,4,5\}}\Sl_{\pmb{\alpha}\in\mathsf{Z}\{2|\sigma(3),\sigma(4),\sigma(5)\}}S[\pmb{\alpha}|2,\sigma(3),\sigma(4),\sigma(5)]A(1,\pmb{\alpha},6).
\eea
For each relative order $\pmb{\sigma}\in\text{perms~}\{3,4,5\}$, the sum over $\pmb{\alpha}\in\mathsf{Z}\{2|\sigma(3),\sigma(4),\sigma(5)\}$ means summing over all possible permutations $\pmb{\alpha}\in \{2\}\shuffle\{\sigma(3),\sigma(4),\sigma(5)\}$ with $\alpha^{-1}(2)>\alpha^{-1}(\sigma(3))$. When all possible $\pmb{\sigma}\in\text{perms~}\{3,4,5\}$ are taken into account, according to the relation \eqref{Eq:GenEquiv2-1} with $\mathsf{H}=\{3,4,5\}$, the above equation converts to the sum of all terms with $\pmb{\alpha}\in \{2\}\shuffle\{\sigma(3),\sigma(4),\sigma(5)\}$ s.t. $\alpha^{-1}(2)<\alpha^{-1}(\sigma(3))$ for all $\pmb{\sigma}\in\text{perms~}\{3,4,5\}$, accompanied by a total minus.
Hence, we arrive
\bea
&&(-1)\Sl_{\pmb{\sigma}\in\text{perms~}\{3,4,5\}}T(2|\pmb{\sigma})\\
&=&\Sl_{\pmb{\sigma}\in\text{perms~}\{3,4,5\}}T(2,\pmb{\sigma})\equiv \Sl_{\pmb{\sigma}\in\text{perms~}\{3,4,5\}}S[2,\sigma(3),\sigma(4),\sigma(5)|2,\sigma(3),\sigma(4),\sigma(5)]A(1,2,\sigma(3),\sigma(4),\sigma(5),6).\nonumber
\eea
The cases $\rho(2)=3,4,5$ are obtained similarly. Finally, \eqref{Eq:DFCMSExampleT2} becomes
\bea
&&\Sl_{\pmb{\sigma}\in\text{perms~}\{3,4,5\}}T(2,\pmb{\sigma})+\Sl_{\pmb{\sigma}\in\text{perms~}\{2,4,5\}}T(3,\pmb{\sigma})+\Sl_{\pmb{\sigma}\in\text{perms~}\{2,3,5\}}T(4,\pmb{\sigma})+\Sl_{\pmb{\sigma}\in\text{perms~}\{2,3,4\}}T(5,\pmb{\sigma})\nn
&=&\Sl_{\pmb{\sigma}\in S_4}S[\pmb{\sigma}|\pmb{\sigma}]A(1,\pmb{\sigma},6),
\eea
which is the RHS of the equivalence condition \eqref{Eq:EquivDFCMS} for $n=6$.

To summarize the above steps, the six-point example for \eqref{Eq:EquivDFCMS} is proved by
\bea
\Bigl[\text{LHS of \eqref{Eq:EquivDFCMS} (for $n=6$)}\Bigr]&=&\Sl_{\pmb{\rho}\in S_4}\,T(\rho(2)|\rho(3)|\rho(4)|\rho(5))\,\,\,=(-1)\Sl_{\pmb{\rho}\in S_4}\,T(\rho(2)|\rho(3),\rho(4),\rho(5))\nn
&=&\Sl_{\pmb{\rho}\in S_4}\,T(\rho(2),\rho(3),\rho(4),\rho(5))=\Bigl[\text{RHS of \eqref{Eq:EquivDFCMS} (for $n=6$)}\Bigr].\Label{Eq:EquivDFCMS6Pattern}
\eea
%
\subsection{General proof of the relation \eqref{Eq:EquivDFCMS}}
Now let us extend the six-point example to a general proof of \eqref{Eq:EquivDFCMS}. As in six-point example, we introduce zigzag permutations for any given $\pmb{\rho}\in S_{n-2}$ by
\bea
&&~~\mathsf{Z}\{\rho(2)|\rho(3)|\dots|\rho(2j)|\rho(2j+1),\dots,\rho(n-1)\}\Label{Eq:Zigzag}\\
&\equiv&\{\pmb{\sigma}|\pmb{\sigma}\in S_{n-2}, \text{s.t.~} \sigma^{-1}({\rho(n-1)})>\sigma^{-1}({\rho(n-2)})>\dots\sigma^{-1}(\rho(2j+2))>\sigma^{-1}(\rho(2j+1)),\nn
&&~~\sigma^{-1}(\rho(2j+1))<\sigma^{-1}(\rho(2j)), \sigma^{-1}(\rho(2j))>\sigma^{-1}(\rho(2j-1)),\dots,\sigma^{-1}(\rho(3))<\sigma^{-1}(\rho(2))\}\nn
&&~~~~~~~~~~~~~~~~~~~~~~~~~~~~~~~~~~~~~~~~~~~~~~~~~~~~~~~~~~~~~~~~~~~~~~~~~~~~~~~~~~~~~~\text{(for  $j\geq 0$ and even $n$)},\nonumber
\eea
where the $j=0$ case is understood as $\mathsf{Z}\{\rho(2),\rho(3),\dots,\rho(n-1)\}\equiv\pmb{\rho}$. We further define a linear combination of amplitudes
\bea
&&T(\rho(2)|\dots|\rho(2j)|\rho(2j+1),\dots,\rho(n-1))
\equiv\Sl_{\small\pmb{\sigma}\in \mathsf{Z}\{\rho(2)|\dots|\rho(2j)|\rho(2j+1),\dots,\rho(n-1)\}}S[\pmb{\sigma}|\pmb{\rho}]A(1,\pmb{\sigma},n),\Label{Eq:DefT}
\eea
in which, the coefficients are momentum kernels. The six-point example (see \eqref{Eq:EquivDFCMS6Pattern}) implies the following recursive relation between $T(\rho(2)|\dots|\rho(2j)|\rho(2j+1),\dots,\rho(n-1))$:
\bea
&&\,\,\,\,\,\,\,\,\,\,\,\,\,\,\Sl_{\pmb{\rho}\in S_{n-2}}T(\rho(2)|\dots|\rho(2j)|\rho(2j+1),\dots,\rho(n-1))\nn&=&(-1)\Sl_{\pmb{\rho}\in S_{n-2}}\,T(\rho(2)|\dots|\rho(2j-2)|\rho(2j-1),\dots,\rho(n-1))\Label{Eq:CombinationT1}~~~~(0\leq j\leq {n-2\over 2}).
\eea
The proof of \eqref{Eq:CombinationT1} is provided in appendix \ref{App:T}. We consider two boundaries of this relation:
\begin{itemize}
\item [(i)] The upper boundary is $j={n-2\over 2}$, for which the LHS of \eqref{Eq:CombinationT1} is {${\sum_{\pmb{\rho}\in S_{n-2}}}T(\rho(2)|\rho(3)|\dots|\rho(n-1))$}. In appendix \ref{Sec:Appendix}, we show that the collection of all $\pmb{\sigma}$ corresponding to a same $\pmb{\rho}$ on the LHS of  \eqref{Eq:EquivDFCMS} is  $\mathsf{Z}\{\rho(2)|\rho(3)|\dots|\rho(n-1)\}$ ({\it i.e.}, $j={n-2\over 2}$). Thus the LHS of \eqref{Eq:CombinationT1} for $j={n-2\over 2}$ is
    \bea
    {\Sl_{\pmb{\rho}\in S_{n-2}}}T(\rho(2)|\rho(3)|\dots|\rho(n-1))=\Sl_{\pmb{\sigma}\in S_{n-2}}\Sl_{\pmb{\rho}\in \mathsf{\Gamma}}S[\pmb{\sigma}|\pmb{\rho}]A(1,\pmb{\sigma},n),
    \eea
    which is the LHS of the equivalence condition \eqref{Eq:EquivDFCMS}.
\item [(ii)] The lower boundary is $j=0$. In this case, the sum on the RHS of \eqref{Eq:CombinationT1} is given by
 \bea\Sl_{\pmb{\rho}\in S_{n-2}}T(\rho(2),\rho(3),\dots,\rho(n-1))=\Sl_{\pmb{\sigma}\in S_{n-2}}S[\pmb{\sigma}|\pmb{\sigma}]A(1,\pmb{\sigma},n)\eea
 which is nothing but (upto a factor ${(-1)^{n-2\over 2}}$) the  RHS of \eqref{Eq:EquivDFCMS}.
\end{itemize}
When we start from the upper boundary and apply the relation \eqref{Eq:CombinationT1} by ${n-2 \over 2}$ times, we arrive the lower boundary with the correct factor {$(-1)^{n-2 \over 2}$}.
Thus the equivalence condition \eqref{Eq:EquivDFCMS} is proven.

\section{Conclusions}\label{Sec:Conclusions}
In this paper, we derived highly-nontrivial generalized BCJ relation \eqref{Eq:NewGaugeIDAmp1} by imposing gauge invariance and dimensional reduction on the graphic expansion of EYM amplitudes. Two additional relations  \eqref{Eq:GenEquiv1} and \eqref{Eq:GenEquiv2} expressed by partial momentum kernels are consequent results of the gauge invariance induced relation \eqref{Eq:NewGaugeIDAmp1}. As an application, we proved the equivalence between  amplitudes constructed by three different types of BCJ numerators. Thus the three approaches (Feynman rules, Abelian Z theory and CHY formula) to NLSM amplitudes are equivalent to each other.
This way we prove the CHY formula of NLSM directly  instead of relying on incomplete evidence, like  the enhanced soft behavior  \cite{Cheung:2015ota}.

There are several further directions. (i) First, generalized BCJ relations induced from the gauge invariance of multi-trace amplitudes deserves further consideration.
(ii) {Second, it seems that the CHY-inspired dimensional reduction is not the unique way to reduce the Lorentz invariants to pure Mandelstam variables. Along the line of unifying relation \cite{Cheung:2017ems}, one can also turn the polarizations to momenta. In addition, other formulations of gauge invariance identities were depicted in \cite{Barreiro:2013dpa,Nandan:2016pya,Boels:2016xhc,Chiodaroli:2017ngp,Boels:2017gyc}. Thus it will be interesting to give a more comprehensive understanding of the gauge invariance induced relations by considering \cite{Cheung:2017ems} and \cite{Barreiro:2013dpa,Nandan:2016pya,Boels:2016xhc,Chiodaroli:2017ngp,Boels:2017gyc}\footnote{We thank Rutger Boels
 for helpful comments on this point.}.}  (iii) As we have seen, the gauge invariance induced relations bridge the DF type BCJ numerators  of NLSM amplitudes and   the compact CMS type ones .  Maybe  they will help us to find  compact polynomial BCJ numerators of YM amplitudes   which are independent of any reference ordering from that of DF type.  We know the sum of BCJ numerators of all possible reference orderings satisfy this requirement, but how about  more compact ones?  (iv) Last but not least,  the gauge invariance induced relations should also exists in string theory.  How about their applications in string amplitudes?

\section*{Acknowledgments}
YD would like to acknowledge Jiangsu Ministry of Science and Technology under contract
BK20170410, NSFC under Grant Nos.11105118, 111547310 as well as the "Fundamental Research Funds for the Central Universities".

\appendix

\section{All graphs for {$\mathsf{H}=\{h_1,h_2,h_3\}$}}\label{App:5Pt}

\begin{figure}[!h]
\centering
\includegraphics[width=7in]{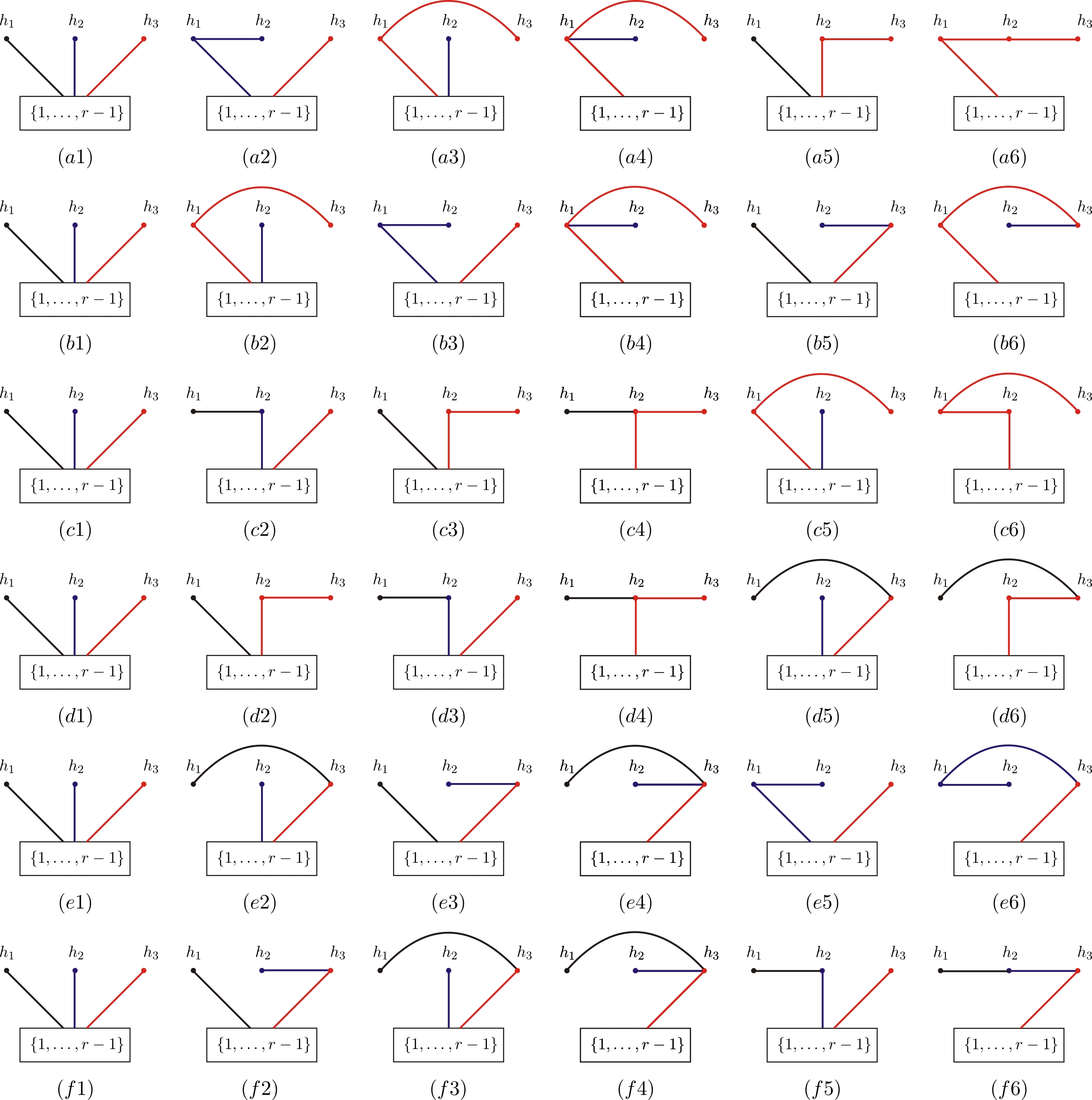}
\caption{All possible graphs with $\mathsf{H}=\{h_1,h_2,h_3\}$.  Graphs in each row contribute to permutations $\{2,\dots,r-1\}\shuffle\,\pmb{\sigma}_{\mathsf{H}}$ for a given relative order $\pmb{\sigma}_{\mathsf{H}}$. }\label{Fig:5ptGraphs}
\end{figure}

When we choose the relative order $R=\{h_1,h_2,h_3\}$, all possible graphs are given by figure \ref{Fig:5ptGraphs}. The correspondence of graphs and the relative permutations $\pmb{\sigma}_{\mathsf{H}}$ is given by
\bea
&&\{h_1,h_2,h_3\}:~~~~(a1)\sim(a6);~~~~\{h_1,h_3,h_2\}:~~~~(b1)\sim (b6);~~~~\{h_2,h_1,h_3\}:~~~~(c1)\sim (c6);\nn
&&\{h_2,h_3,h_1\}:~~~~(d1)\sim(d6);~~~~\{h_3,h_1,h_2\}:~~~~(e1)\sim (e6);~~~~\{h_3,h_2,h_1\}:~~~~(f1)\sim (f6).
\eea

\section{Proof of \eqref{Eq:CombinationT1}}\label{App:T}

To prove the relation \eqref{Eq:CombinationT1}, we consider the LHS for a given $j$:
\bea
\Sl_{\pmb{\rho}\in S_{n-2}}T(\rho(2)|\dots|\rho(2j)|\rho(2j+1),\dots,\rho(n-1)). \Label{Eq:CombinationT}
\eea
Assuming  $\mathsf{I}_{2j-1}\equiv\{i_2,i_3,\dots,i_{2j}\}$  with $2j-1$ elements is a subset of $\{2,\dots,n-1\}$, we can divide the set $\{2,\dots,n-1\}$ into two parts
 $\{i_2,i_3,\dots,i_{2j}\}$ and its complement $\overline{\mathsf{I}_{2j-1}}=\{2,\dots,n-1\}\setminus \{i_2,i_3,\dots,i_{2j}\}$. Then \eqref{Eq:CombinationT} can be arranged as
 \bea
&&\Sl_{\pmb{\rho}\in S_{n-2}}T(\rho(2)|\dots|\rho(2j)|\rho(2j+1),\dots,\rho(n-1))\Label{Eq:DFCMSGenProof0}\nn
&=&\Sl_{\mathsf{I}_{2j-1}\subseteq\{2,\dots,n-1\}}\Bigl[\Sl_{\pmb{\rho}_A\in\text{perms~$\mathsf{I}_{2j-1}$}}\Sl_{\pmb{\rho}_B\in \text{perms~$\overline{\mathsf{I}_{2j-1}}$}}T(\rho_A(2)|\dots|\rho_A(2j)|\rho_B(2j+1),\dots,\rho_B(n-1))\Bigr].
\eea
 in which, the first summation is over all possible choices of the subset $\mathsf{I}_{2j-1}$ for fixed $j$, the second and the third summations are given by summing over all possible permutations of elements in $\mathsf{I}_{2j-1}$ and $\overline{\mathsf{I}_{2j-1}}$.
For given $\mathsf{I}_{2j-1}$ and given $\pmb{\rho}_A\in\text{perms~$\mathsf{I}_{2j-1}$}$, we write the sum over $\pmb{\rho}_B$ explicitly
\bea
&&\Sl_{\pmb{\rho}_B\in \text{perms~$\overline{\mathsf{I}_{2j-1}}$}}T(\rho_A(2)|\dots|\rho_A(2j-2)|\rho_A(2j-1)|\rho_A(2j)|\rho_B(2j+1),\dots,\rho_B(n-1))\nn
&=&\Sl_{\pmb{\rho}_B\in \text{perms~$\overline{\mathsf{I}_{2j-1}}$}}\Sl_{\pmb{\sigma}\in \mathsf{Z}\{\rho_A(2)|\dots|\rho_A(2j)|\pmb{\rho}_B\}}S[\pmb{\sigma}|\pmb{\rho}_A,\pmb{\rho}_B]A(1,\pmb{\sigma},n).\Label{Eq:DFCMSGenProof1}
\eea
According to the definition of zigzag pattern \eqref{Eq:Zigzag}, the sum over $\pmb{\sigma}$ in the above equation can be realized by the following two steps:
(i) first fix a relative order $\pmb{\sigma}_A$ of $\rho_A(2),\rho_A(3),\dots,\rho_A(2j)$, s.t.,
\bea
&&\pmb{\sigma}_A\in \Bigl\{\text{perms~}\pmb{\rho}_A\,\text{s.t.}\,\sigma_A^{-1}(\rho_A(2j-1))<\sigma_A^{-1}(\rho_A(2j)),\sigma_A^{-1}(\rho_A(2j-2))>\sigma_A^{-1}(\rho_A(2j-1)),\nn
&&\dots,\sigma_A^{-1}(\rho_A(3))<\sigma_A^{-1}(\rho_A(2))\Bigr\},\Label{Eq:sigmaA}
\eea
and sum over all possible permutations $\pmb{\sigma}\in\pmb{\sigma}_A\shuffle\pmb{\rho}_B$ s.t. $\sigma^{-1}(\rho_B(2j+1))<\sigma^{-1}(\rho_A(2j))$, (ii) sum over all possible $\pmb{\sigma}_A$ satisfying \eqref{Eq:sigmaA}. Since $\pmb{\rho}_B$ and $\pmb{\sigma}_A$ are permutations of elements from two disjointed sets, {the sums over them} commute with each other.
Therefore,  \eqref{Eq:DFCMSGenProof1} becomes
\bea
\Sl_{\pmb{\sigma}_A}\biggl[\Sl_{\pmb{\rho}_B\in \text{perms~$\overline{\mathsf{I}_{2j-1}}$}}\Sl_{\substack{\pmb{\sigma}\in\pmb{\sigma}_A\shuffle\,\pmb{\rho}_B,~\text{s.t.}\\ \sigma^{-1}(\rho_B(2j+1))<\sigma^{-1}(\rho_A(2j))}}S[\pmb{\sigma}|\pmb{\rho}_A,\pmb{\rho}_B]A(1,\pmb{\sigma},n)\biggr],
\eea
where the first summation is taken over all $\pmb{\sigma}_A$ satisfying \eqref{Eq:sigmaA}. For a given $\pmb{\sigma}_A$ satisfying \eqref{Eq:sigmaA}, one can apply the relation \eqref{Eq:GenEquiv2-1} to the expression in the square brackets. Thus the above expression evaluates to
\bea&&(-1)\Sl_{\pmb{\rho}_B\in \text{perms~$\overline{\mathsf{I}_{2j-1}}$}}\biggl[\Sl_{\pmb{\sigma}_A}\Sl_{\substack{\pmb{\sigma}\in\pmb{\sigma}_A\shuffle\,\pmb{\rho}_B,~\text{s.t.}\\ \sigma^{-1}(\rho_A(2j))<\sigma^{-1}(\rho_B(2j+1))}}S[\pmb{\sigma}|\pmb{\rho}_A,\pmb{\rho}_B]A(1,\pmb{\sigma},n)\biggr]\nn
&=&(-1)\Sl_{\pmb{\rho}_B\in \text{perms~$\overline{\mathsf{I}_{2j-1}}$}}\biggl[\Sl_{\pmb{\sigma}\in Z\{\rho_A(2)|\dots|\rho_A(2j-1),\rho_A(2j),\,\pmb{\rho}_B \}}S[\pmb{\sigma}|\pmb{\rho}_A,\pmb{\rho}_B]A(1,\pmb{\sigma},n)\biggr]\nn
&=&(-1)\Sl_{\pmb{\rho}_B\in \text{perms~$\overline{\mathsf{I}_{2j-1}}$}}T(\rho_A(2)|\dots|\rho_A(2j-2)|\rho_A(2j-1),\,\rho_A(2j),\,\pmb{\rho}_B),
\eea
in which the second equality is obtained by considering the definition of zigzag permutations \eqref{Eq:Zigzag}:
\bea
 &&Z\{\rho_A(2)|\dots|\rho_A(2j-1),\rho_A(2j),\,\pmb{\rho}_B \}\\
 &=&\{\pmb{\sigma}|\pmb{\sigma}\in S_{n-2}, \text{s.t.~} \sigma^{-1}({\rho_B(n-1)})>\dots>\sigma^{-1}(\rho_B(2j+1))>\sigma^{-1}(\rho_A(2j))>\sigma^{-1}(\rho_A(2j-1)),\nn
&&\sigma^{-1}(\rho_A(2j-2))>\sigma^{-1}(\rho_A(2j-1)), \sigma^{-1}(\rho_A(2j-3))<\sigma^{-1}(\rho_A(2j-2)),\dots,\sigma^{-1}(\rho_A(3))<\sigma^{-1}(\rho_A(2))\}\nn
&=&\{\pmb{\sigma}|\pmb{\sigma}\in\pmb{\sigma}_A\shuffle\,\pmb{\rho}_B, \text{s.t\,} \sigma_A^{-1}(\rho_A(2j-1))<\sigma_A^{-1}(\rho_A(2j)),\sigma_A^{-1}(\rho_A(2j-2))>\sigma_A^{-1}(\rho_A(2j-1)),\nn
&&\dots,\sigma_A^{-1}(\rho_A(3))<\sigma_A^{-1}(\rho_A(2))\text{\,and\,}\sigma^{-1}(\rho_A(2j))<\sigma^{-1}(\rho_B(2j+1))\}.\nonumber
\eea
Consequently, \eqref{Eq:DFCMSGenProof0} becomes
\bea
&&\Sl_{\pmb{\rho}\in S_{n-2}}T(\rho(2)|\dots|\rho(2j)|\rho(2j+1),\dots,\rho(n-1))\Label{Eq:DFCMSGenProof2}\\
&=&(-1)\Sl_{\mathsf{I}_{2j-1}\subseteq\{2,\dots,n-1\}}\Bigl[\Sl_{\pmb{\rho}_A\in\text{perms~$\mathsf{I}_{2j-1}$}}\Sl_{\pmb{\rho}_B\in \text{perms~$\overline{\mathsf{I}_{2j-1}}$}}T(\rho_A(2)|\dots|\rho_A(2j-2)|\rho_A(2j-1),\,\rho_A(2j),\,\pmb{\rho}_B)\Bigr].\nonumber
\eea

Now let us understand the summations on the RHS of  \eqref{Eq:DFCMSGenProof2}. Given $\mathsf{I}_{2j-1}$, we collect terms with  $\rho_A(2j-1)=a$, $\rho_A(2j)=b$, (for given $a,b,\in \mathsf{I}_{2j-1}$) then obtain a term
\bea
\Sl_{\pmb{\rho}_{A'}\in\text{perms~$\mathsf{I}_{2j-3}$}}\Sl_{\pmb{\rho}_B\in \text{perms~$\overline{\mathsf{I}_{2j-1}}$}}T(\rho_{A'}(2)|\dots|\rho_{A'}(2j-2)|a,\,b,\,\pmb{\rho}_B),
\eea
where we define $\mathsf{I}_{2j-3}\equiv\mathsf{I}_{2j-1}\setminus \{a,b\}$.
Correspondingly, we also have other terms in $\eqref{Eq:DFCMSGenProof2}$ with distinct  $\mathsf{I}_{2j-1}$  (identical $\mathsf{I}_{2j-1}$ for the special case with $\rho_A(2j-1)=b$, $\rho_A(2j)=a$) but a same $\mathsf{I}_{2j-3}\equiv\mathsf{I}_{2j-1}\setminus \{x,y\}$, where $\rho_A(2j-1)=x$, $\rho_A(2j)=y$ for an ordered pair $(x,y)$ satisfying $x,y\in\{a,b\}\cup\overline{\mathsf{I}_{2j-1}}=\overline{\mathsf{I}_{2j-3}}$. The sum of all such terms gives rise
\bea
&&\Sl_{\pmb{\rho}_{A'}\in\text{perms~$\mathsf{I}_{2j-3}$}}\,\,\biggl[\Sl_{x,y\in\overline{\mathsf{I}_{2j-3}}}\,\,\Sl_{\{\rho_A(2j-1),\rho_A(2j)\}\in\text{perms }\{x,y\}}\,\,\Sl_{\pmb{\rho}_B\in \text{perms~$\overline{\mathsf{I}_{2j-3}}\setminus \{x,y\}$}}\nn
&&~~~~~~~~~~~~~~~~~~~~~~~~~~~~~~~~~~~~~~~~~~~~~~~~~~~~~~~~~~T(\rho_{A'}(2)|\dots|\rho_{A'}(2j-2)|\rho_A(2j-1),\rho_A(2j),\pmb{\rho}_B)\biggr].
\eea
Defining $\rho_A(2j-1)\equiv\rho_{B'}(2j-1)$, $\rho_A(2j)\equiv\rho_{B'}(2j)$, $\rho_B(2j+1)\equiv\rho_{B'}(2j+1)$, ..., $\rho_B(n-1)\equiv\rho_{B'}(n-1)$ and noting that for given $\pmb{\rho}_{A'}\in\text{perms~$\mathsf{I}_{2j-3}$}$ the other three summations becomes $\sum_{\pmb{\rho}_{B'}\in \text{perms~}\overline{\mathsf{I}_{2j-3}}}$, we reformulate the above expression as
\bea
\Sl_{\pmb{\rho}_{A'}\in\text{perms~$\mathsf{I}_{2j-3}$}}\,\Sl_{\pmb{\rho}_{B'}\in \text{perms~}\overline{\mathsf{I}_{2j-3}}}\,T(\rho_{A'}(2)|\dots|\rho_{A'}(2j-2)|\pmb{\rho}_{B'}).
\eea
Summing over all possible choices of $\mathsf{I}_{2j-3}\subseteq\{2,\dots,n-1\}$, we finally express the RHS of \eqref{Eq:DFCMSGenProof2} by
\bea
&&(-1)\Sl_{\mathsf{I}_{2j-3}\subseteq\{2,\dots,n-1\}}\,\Sl_{\pmb{\rho}_{A'}\in\text{perms~$\mathsf{I}_{2j-3}$}}\,\Sl_{\pmb{\rho}_{B'}\in \text{perms~}\overline{\mathsf{I}_{2j-3}}}\,T(\rho_{A'}(2)|\dots|\rho_{A'}(2j-2)|\pmb{\rho}_{B'})\nn
&=&(-1)\Sl_{\pmb{\rho}\in S_{n-2}}\,T(\rho(2)|\dots|\rho(2j-2)|\rho(2j-1),\rho(2j),\dots,\rho(n)).
\eea
Hence the relation \eqref{Eq:CombinationT} is proven.

\section{Understanding the zigzag pattern of $\pmb{\sigma}$  for given $\pmb{\rho}$ in \eqref{Eq:EquivDFCMS} }\label{Sec:Appendix}
We first show that, if a given $\pmb{\rho}=\{\rho(2),\rho(3),\dots,\rho(n-1)\}$ on the LHS of \eqref{Eq:EquivDFCMS} can be considered as {a permutation in} $\mathsf{\Gamma}(\pmb{\sigma})$ for some permutation $\pmb{\sigma}$, the $\pmb{\sigma}$ must satisfy the zigzag pattern, {\it i.e.}, $\pmb{\sigma}\in \mathsf{Z}\{\rho(2)|\rho(3)|\dots|\rho(n-1)\}$. This can be understood as follows:
\begin{itemize}
\item  As defined in section \ref{Sec:NumeratorsNLSM}, the element $n$ is always considered as the last element in both $\pmb{\sigma}$ and $\pmb{\rho}$, thus we have $\sigma^{-1}(\rho(n-1))<\sigma^{-1}(n)$.  Since there is no element between $\rho(n-1)$ and $n$ in the permutation $\pmb{\rho}$, according to the rule given in section \ref{Sec:NumeratorsNLSM} (see the point (ii) below \eqref{Eq:DFform}), we deduce $\sigma^{-1}(\rho(n-2))>\sigma^{-1}(\rho(n-1))$;

\item We now consider $\rho(n-2)$. In the  permutation $\pmb{\rho}$, there is one element $\rho(n-1)$, which satisfies  $\sigma^{-1}(\rho(n-1))<\sigma^{-1}(\rho(n-2))$, between $\rho(n-2)$ and $n$ (note that $\sigma^{-1}(\rho(n-2))<\sigma^{-1}(n)$). According to the rule given in section \ref{Sec:NumeratorsNLSM} (see the point (i) below \eqref{Eq:DFform}), we deduce that $\sigma^{-1}(\rho(n-3))<\sigma^{-1}(\rho(n-2))$.

\item We further consider $\rho(n-3)$. Since $\sigma^{-1}(\rho(n-3))<\sigma^{-1}(\rho(n-2))$ and there is no element between $\rho(n-3)$ and $\rho(n-2)$ in the permutation $\pmb{\rho}$, we must have $\sigma^{-1}(\rho(n-4))>\sigma^{-1}(\rho(n-3))$, in accordance to the point (ii) below \eqref{Eq:DFform}.

\item We turn to $\rho(n-4)$. Since  $\sigma^{-1}(\rho(n-4))>\sigma^{-1}(\rho(n-3))$, we should have $\sigma^{-1}(\rho(n-5))<\sigma^{-1}(\rho(n-4))$ due to the point (i) below \eqref{Eq:DFform}).

\item Repeat the above discussions, we find the general condition
 \bea\sigma^{-1}(\rho(2j+2))>\sigma^{-1}(\rho(2j+1)),~~~~\sigma^{-1}(\rho(2j+1))<\sigma^{-1}(\rho(2j)),~~~~(\text{for $j\geq1$}).\eea
    Thus the permutation $\pmb{\sigma}$ must be in $\mathsf{Z}\{\rho(2)|\rho(3)|\dots|\rho(n-1)\}$ for given $\pmb{\rho}$.
\end{itemize}
Conversely, we show that  $\pmb{\rho}$ must be in $\mathsf{{\Gamma}}(\pmb{\sigma})$ for any permutation $\pmb{\sigma}\in \mathsf{Z}\{\rho(2)|\rho(3)|\dots|\rho(n-1)\}$. This is because:
\begin{itemize}
\item For any $\pmb{\sigma}\in \mathsf{Z}\{\rho(2)|\rho(3)|\dots|\rho(n-1)\}$, if $\sigma(a)=\rho(2j+1)$ ($\sigma(a)\in\pmb{\sigma}$), we must have some $b>a$ and $c>a$ s.t. $\sigma(b)=\rho(2j+2)$ and $\sigma(c)=\rho(2j)$. In the permutation $\pmb{\rho}$, $\rho(2j+2)$ and $\rho(2j)$ are the nearest elements on the RHS and LHS satisfying
$a=\sigma^{-1}(2j+1)<b=\sigma^{-1}(2j)$ and $a=\sigma^{-1}(\rho(2j+1))<c=\sigma^{-1}(\rho(2j+2))$. In addition, there is no element between $\rho(2j+2)$, $\rho(2j+1)$ and $\rho(2j)$, $\rho(2j+1)$ in the permutation $\pmb{\rho}$. Thus the condition (ii) below \eqref{Eq:DFform} in section \ref{Sec:NumeratorsNLSM} is satisfied.
\item For any $\pmb{\sigma}\in \mathsf{Z}\{\rho(2)|\rho(3)|\dots|\rho(n-1)\}$, if $\sigma(a)=\rho(2j)$ ($\sigma(a)\in\pmb{\sigma}$), we have two possibilities.
\begin{itemize}
\item If the nearest element $\sigma(b)$ (and  $\sigma(c)$) on the LSH (and RHS) to $\rho(2j)$ in permutation $\pmb{\rho}$ s.t.  $b>a$ (and $c>a$) has the form $\sigma(b)=\rho(2k)$ (and $\sigma(c)=\rho(2k')$), we must have odd number of elements between $\sigma(b)=\rho(2k)$ (and $\sigma(c)=\rho(2k')$) and $\sigma(a)=\rho(2j)$ in $\pmb{\rho}$ (because there must be odd numbers between two even numbers). Thus the condition (i) in section \ref{Sec:NumeratorsNLSM} is satisfied;
\item Assuming the nearest element $\sigma(b)$ (or $\sigma(c)$) on the LSH (or RHS) to $\rho(2j)$ in  $\pmb{\rho}$, which satisfies $b>a$ (or $c>a$) has the form $\sigma(b)=\rho(2k+1)$ (or $\sigma(c)=\rho(2k'+1)$), we always have $\rho(2k+2)$ (or $\rho(2k')$), which is more nearer to $\sigma(a)=\rho(2j)$ in $\pmb{\rho}$ and satisfies $\sigma^{-1}(\rho(2k+2))>\sigma^{-1}(\rho(2k+1))=\sigma(b)>\sigma(a)=\rho(2j)$ (or $\sigma^{-1}(\rho(2k'))>\sigma^{-1}(\rho(2k'+1))=\sigma(c)>\sigma(a)=\rho(2j)$). Thus we return to the previous case.
\end{itemize}
\end{itemize}

\bibliographystyle{JHEP}
\bibliography{GaugeInvAndNumeratorsInNLSM}

\providecommand{\href}[2]{#2}\begingroup\raggedright\begin{thebibliography}{10}

\bibitem{Bern:2008qj}
Z.~Bern, J.~J.~M. Carrasco, and H.~Johansson, {\it {New Relations for
  Gauge-Theory Amplitudes}},  {\em Phys. Rev.} {\bf D78} (2008) 085011,
  [\href{http://arxiv.org/abs/0805.3993}{{\tt arXiv:0805.3993}}].

\bibitem{Bern:2010ue}
Z.~Bern, J.~J.~M. Carrasco, and H.~Johansson, {\it {Perturbative Quantum
  Gravity as a Double Copy of Gauge Theory}},  {\em Phys. Rev. Lett.} {\bf 105}
  (2010) 061602, [\href{http://arxiv.org/abs/1004.0476}{{\tt
  arXiv:1004.0476}}].

\bibitem{Kleiss:1988ne}
R.~Kleiss and H.~Kuijf, {\it {Multi - Gluon Cross-sections and Five Jet
  Production at Hadron Colliders}},  {\em Nucl. Phys.} {\bf B312} (1989)
  616--644.

\bibitem{Feng:2010my}
B.~Feng, R.~Huang, and Y.~Jia, {\it {Gauge Amplitude Identities by On-shell
  Recursion Relation in S-matrix Program}},  {\em Phys. Lett.} {\bf B695}
  (2011) 350--353, [\href{http://arxiv.org/abs/1004.3417}{{\tt
  arXiv:1004.3417}}].

\bibitem{Chen:2011jxa}
Y.-X. Chen, Y.-J. Du, and B.~Feng, {\it {A Proof of the Explicit Minimal-basis
  Expansion of Tree Amplitudes in Gauge Field Theory}},  {\em JHEP} {\bf 02}
  (2011) 112, [\href{http://arxiv.org/abs/1101.0009}{{\tt arXiv:1101.0009}}].

\bibitem{BjerrumBohr:2009rd}
N.~E.~J. Bjerrum-Bohr, P.~H. Damgaard, and P.~Vanhove, {\it {Minimal Basis for
  Gauge Theory Amplitudes}},  {\em Phys. Rev. Lett.} {\bf 103} (2009) 161602,
  [\href{http://arxiv.org/abs/0907.1425}{{\tt arXiv:0907.1425}}].

\bibitem{Stieberger:2009hq}
S.~Stieberger, {\it {Open \& Closed vs. Pure Open String Disk Amplitudes}},
  \href{http://arxiv.org/abs/0907.2211}{{\tt arXiv:0907.2211}}.

\bibitem{Chen:2013fya}
G.~Chen and Y.-J. Du, {\it {Amplitude Relations in Non-linear Sigma Model}},
  {\em JHEP} {\bf 01} (2014) 061, [\href{http://arxiv.org/abs/1311.1133}{{\tt
  arXiv:1311.1133}}].

\bibitem{Cachazo:2013gna}
F.~Cachazo, S.~He, and E.~Y. Yuan, {\it {Scattering equations and
  Kawai-Lewellen-Tye orthogonality}},  {\em Phys. Rev.} {\bf D90} (2014), no.~6
  065001, [\href{http://arxiv.org/abs/1306.6575}{{\tt arXiv:1306.6575}}].

\bibitem{Cachazo:2013hca}
F.~Cachazo, S.~He, and E.~Y. Yuan, {\it {Scattering of Massless Particles in
  Arbitrary Dimensions}},  {\em Phys. Rev. Lett.} {\bf 113} (2014), no.~17
  171601, [\href{http://arxiv.org/abs/1307.2199}{{\tt arXiv:1307.2199}}].

\bibitem{Cachazo:2013iea}
F.~Cachazo, S.~He, and E.~Y. Yuan, {\it {Scattering of Massless Particles:
  Scalars, Gluons and Gravitons}},  {\em JHEP} {\bf 07} (2014) 033,
  [\href{http://arxiv.org/abs/1309.0885}{{\tt arXiv:1309.0885}}].

\bibitem{Cachazo:2014xea}
F.~Cachazo, S.~He, and E.~Y. Yuan, {\it {Scattering Equations and Matrices:
  From Einstein To Yang-Mills, DBI and NLSM}},  {\em JHEP} {\bf 07} (2015) 149,
  [\href{http://arxiv.org/abs/1412.3479}{{\tt arXiv:1412.3479}}].

\bibitem{Ma:2011um}
Q.~Ma, Y.-J. Du, and Y.-X. Chen, {\it {On Primary Relations at Tree-level in
  String Theory and Field Theory}},  {\em JHEP} {\bf 02} (2012) 061,
  [\href{http://arxiv.org/abs/1109.0685}{{\tt arXiv:1109.0685}}].

\bibitem{Britto:2004ap}
R.~Britto, F.~Cachazo, and B.~Feng, {\it {New recursion relations for tree
  amplitudes of gluons}},  {\em Nucl. Phys.} {\bf B715} (2005) 499--522,
  [\href{http://arxiv.org/abs/hep-th/0412308}{{\tt hep-th/0412308}}].

\bibitem{Britto:2005fq}
R.~Britto, F.~Cachazo, B.~Feng, and E.~Witten, {\it {Direct proof of tree-level
  recursion relation in Yang-Mills theory}},  {\em Phys. Rev. Lett.} {\bf 94}
  (2005) 181602, [\href{http://arxiv.org/abs/hep-th/0501052}{{\tt
  hep-th/0501052}}].

\bibitem{Du:2016tbc}
Y.-J. Du and C.-H. Fu, {\it {Explicit BCJ numerators of nonlinear simga
  model}},  {\em JHEP} {\bf 09} (2016) 174,
  [\href{http://arxiv.org/abs/1606.05846}{{\tt arXiv:1606.05846}}].

\bibitem{Kawai:1985xq}
H.~Kawai, D.~C. Lewellen, and S.~H.~H. Tye, {\it {A Relation Between Tree
  Amplitudes of Closed and Open Strings}},  {\em Nucl. Phys.} {\bf B269} (1986)
  1.

\bibitem{Bern:1998ug}
Z.~Bern, L.~J. Dixon, D.~C. Dunbar, M.~Perelstein, and J.~S. Rozowsky, {\it {On
  the relationship between Yang-Mills theory and gravity and its implication
  for ultraviolet divergences}},  {\em Nucl. Phys.} {\bf B530} (1998) 401--456,
  [\href{http://arxiv.org/abs/hep-th/9802162}{{\tt hep-th/9802162}}].

\bibitem{BjerrumBohr:2010ta}
N.~E.~J. Bjerrum-Bohr, P.~H. Damgaard, B.~Feng, and T.~Sondergaard, {\it
  {Gravity and Yang-Mills Amplitude Relations}},  {\em Phys. Rev.} {\bf D82}
  (2010) 107702, [\href{http://arxiv.org/abs/1005.4367}{{\tt
  arXiv:1005.4367}}].

\bibitem{BjerrumBohr:2010zb}
N.~E.~J. Bjerrum-Bohr, P.~H. Damgaard, B.~Feng, and T.~Sondergaard, {\it {New
  Identities among Gauge Theory Amplitudes}},  {\em Phys. Lett.} {\bf B691}
  (2010) 268--273, [\href{http://arxiv.org/abs/1006.3214}{{\tt
  arXiv:1006.3214}}].

\bibitem{BjerrumBohr:2010yc}
N.~E.~J. Bjerrum-Bohr, P.~H. Damgaard, B.~Feng, and T.~Sondergaard, {\it {Proof
  of Gravity and Yang-Mills Amplitude Relations}},  {\em JHEP} {\bf 09} (2010)
  067, [\href{http://arxiv.org/abs/1007.3111}{{\tt arXiv:1007.3111}}].

\bibitem{BjerrumBohr:2010hn}
N.~E.~J. Bjerrum-Bohr, P.~H. Damgaard, T.~Sondergaard, and P.~Vanhove, {\it
  {The Momentum Kernel of Gauge and Gravity Theories}},  {\em JHEP} {\bf 01}
  (2011) 001, [\href{http://arxiv.org/abs/1010.3933}{{\tt arXiv:1010.3933}}].

\bibitem{Carrasco:2016ldy}
J.~J.~M. Carrasco, C.~R. Mafra, and O.~Schlotterer, {\it {Abelian Z-theory:
  NLSM amplitudes and $\alpha$'-corrections from the open string}},  {\em JHEP}
  {\bf 06} (2017) 093, [\href{http://arxiv.org/abs/1608.02569}{{\tt
  arXiv:1608.02569}}].

\bibitem{Du:2017kpo}
Y.-J. Du and F.~Teng, {\it {BCJ numerators from reduced Pfaffian}},  {\em JHEP}
  {\bf 04} (2017) 033, [\href{http://arxiv.org/abs/1703.05717}{{\tt
  arXiv:1703.05717}}].

\bibitem{Stieberger:2016lng}
S.~Stieberger and T.~R. Taylor, {\it {New relations for Einstein-Yang-Mills
  amplitudes}},  {\em Nucl. Phys.} {\bf B913} (2016) 151--162,
  [\href{http://arxiv.org/abs/1606.09616}{{\tt arXiv:1606.09616}}].

\bibitem{Nandan:2016pya}
D.~Nandan, J.~Plefka, O.~Schlotterer, and C.~Wen, {\it {Einstein-Yang-Mills
  from pure Yang-Mills amplitudes}},  {\em JHEP} {\bf 10} (2016) 070,
  [\href{http://arxiv.org/abs/1607.05701}{{\tt arXiv:1607.05701}}].

\bibitem{delaCruz:2016gnm}
L.~de~la Cruz, A.~Kniss, and S.~Weinzierl, {\it {Relations for
  Einstein-Yang-Mills amplitudes from the CHY representation}},  {\em Phys.
  Lett.} {\bf B767} (2017) 86--90, [\href{http://arxiv.org/abs/1607.06036}{{\tt
  arXiv:1607.06036}}].

\bibitem{Schlotterer:2016cxa}
O.~Schlotterer, {\it {Amplitude relations in heterotic string theory and
  Einstein-Yang-Mills}},  {\em JHEP} {\bf 11} (2016) 074,
  [\href{http://arxiv.org/abs/1608.00130}{{\tt arXiv:1608.00130}}].

\bibitem{Fu:2017uzt}
C.-H. Fu, Y.-J. Du, R.~Huang, and B.~Feng, {\it {Expansion of
  Einstein-Yang-Mills Amplitude}},  {\em JHEP} {\bf 09} (2017) 021,
  [\href{http://arxiv.org/abs/1702.08158}{{\tt arXiv:1702.08158}}].

\bibitem{Chiodaroli:2017ngp}
M.~Chiodaroli, M.~Gunaydin, H.~Johansson, and R.~Roiban, {\it {Explicit
  Formulae for Yang-Mills-Einstein Amplitudes from the Double Copy}},  {\em
  JHEP} {\bf 07} (2017) 002, [\href{http://arxiv.org/abs/1703.00421}{{\tt
  arXiv:1703.00421}}].

\bibitem{Teng:2017tbo}
F.~Teng and B.~Feng, {\it {Expanding Einstein-Yang-Mills by Yang-Mills in CHY
  frame}},  {\em JHEP} {\bf 05} (2017) 075,
  [\href{http://arxiv.org/abs/1703.01269}{{\tt arXiv:1703.01269}}].

\bibitem{Du:2017gnh}
Y.-J. Du, B.~Feng, and F.~Teng, {\it {Expansion of All Multitrace Tree Level
  EYM Amplitudes}},  \href{http://arxiv.org/abs/1708.04514}{{\tt
  arXiv:1708.04514}}.

\bibitem{DelDuca:1999rs}
V.~Del~Duca, L.~J. Dixon, and F.~Maltoni, {\it New color decompositions for
  gauge amplitudes at tree and loop level},  {\em Nucl. Phys. B} {\bf 571}
  (2000) 51--70, [\href{http://arxiv.org/abs/hep-ph/9910563}{{\tt
  hep-ph/9910563}}].

\bibitem{Kiermaier}
M.~Kiermaier, {\it talk at amplitudes 2010, may 2010 at qmul, london, uk.
  http://www.strings.ph.qmul.ac.uk/\textasciitilde{}theory/amplitudes2010/talks/mk2010.pdf},
  .

\bibitem{Bern:2010yg}
Z.~Bern, T.~Dennen, Y.-t. Huang, and M.~Kiermaier, {\it {Gravity as the Square
  of Gauge Theory}},  {\em Phys. Rev.} {\bf D82} (2010) 065003,
  [\href{http://arxiv.org/abs/1004.0693}{{\tt arXiv:1004.0693}}].

\bibitem{Mafra:2011kj}
C.~R. Mafra, O.~Schlotterer, and S.~Stieberger, {\it {Explicit BCJ Numerators
  from Pure Spinors}},  {\em JHEP} {\bf 07} (2011) 092,
  [\href{http://arxiv.org/abs/1104.5224}{{\tt arXiv:1104.5224}}].

\bibitem{Du:2011js}
Y.-J. Du, B.~Feng, and C.-H. Fu, {\it {BCJ Relation of Color Scalar Theory and
  KLT Relation of Gauge Theory}},  {\em JHEP} {\bf 08} (2011) 129,
  [\href{http://arxiv.org/abs/1105.3503}{{\tt arXiv:1105.3503}}].

\bibitem{Fu:2012uy}
C.-H. Fu, Y.-J. Du, and B.~Feng, {\it {An algebraic approach to BCJ
  numerators}},  {\em JHEP} {\bf 03} (2013) 050,
  [\href{http://arxiv.org/abs/1212.6168}{{\tt arXiv:1212.6168}}].

\bibitem{Fu:2013qna}
C.-H. Fu, Y.-J. Du, and B.~Feng, {\it {Note on Construction of Dual-trace
  Factor in Yang-Mills Theory}},  {\em JHEP} {\bf 10} (2013) 069,
  [\href{http://arxiv.org/abs/1305.2996}{{\tt arXiv:1305.2996}}].

\bibitem{Du:2013sha}
Y.-J. Du, B.~Feng, and C.-H. Fu, {\it {The Construction of Dual-trace Factor in
  Yang-Mills Theory}},  {\em JHEP} {\bf 07} (2013) 057,
  [\href{http://arxiv.org/abs/1304.2978}{{\tt arXiv:1304.2978}}].

\bibitem{Fu:2014pya}
C.-H. Fu, Y.-J. Du, and B.~Feng, {\it {Note on symmetric BCJ numerator}},  {\em
  JHEP} {\bf 08} (2014) 098, [\href{http://arxiv.org/abs/1403.6262}{{\tt
  arXiv:1403.6262}}].

\bibitem{Cachazo:2014nsa}
F.~Cachazo, S.~He, and E.~Y. Yuan, {\it {Einstein-Yang-Mills Scattering
  Amplitudes From Scattering Equations}},  {\em JHEP} {\bf 01} (2015) 121,
  [\href{http://arxiv.org/abs/1409.8256}{{\tt arXiv:1409.8256}}].

\bibitem{Barreiro:2013dpa}
L.~A. Barreiro and R.~Medina, {\it {RNS derivation of N-point disk amplitudes
  from the revisited S-matrix approach}},  {\em Nucl. Phys.} {\bf B886} (2014)
  870--951, [\href{http://arxiv.org/abs/1310.5942}{{\tt arXiv:1310.5942}}].

\bibitem{Boels:2016xhc}
R.~H. Boels and R.~Medina, {\it {Graviton and gluon scattering from first
  principles}},  {\em Phys. Rev. Lett.} {\bf 118} (2017) 061602,
  [\href{http://arxiv.org/abs/1607.08246}{{\tt arXiv:1607.08246}}].

\bibitem{Boels:2017gyc}
R.~H. Boels and H.~Luo, {\it {A minimal approach to the scattering of physical
  massless bosons}},  \href{http://arxiv.org/abs/1710.10208}{{\tt
  arXiv:1710.10208}}.

\bibitem{Cheung:2015ota}
C.~Cheung, K.~Kampf, J.~Novotny, C.-H. Shen, and J.~Trnka, {\it {On-Shell
  Recursion Relations for Effective Field Theories}},  {\em Phys. Rev. Lett.}
  {\bf 116} (2016), no.~4 041601, [\href{http://arxiv.org/abs/1509.03309}{{\tt
  arXiv:1509.03309}}].

\bibitem{Cheung:2017ems}
C.~Cheung, C.-H. Shen, and C.~Wen, {\it {Unifying Relations for Scattering
  Amplitudes}},  {\em JHEP} {\bf 02} (2018) 095,
  [\href{http://arxiv.org/abs/1705.03025}{{\tt arXiv:1705.03025}}].

\end{thebibliography}\endgroup

\end{document}